\documentclass[twocolumn,trackchanges]{aastex701}

\usepackage{makecell}
\usepackage{ragged2e}
\usepackage{threeparttable}
\usepackage{multirow}
\usepackage[utf8]{inputenc}
\usepackage[T1]{fontenc}
\usepackage{amsmath} 

\newcommand{\psr}{PSR\,J1631$-$4722}
\newcommand{\pwn}{J1631$-$4721}
\newcommand{\snr}{G336.7+0.5}
\newcommand{\pwnname}{Thunder}
\newcommand{\snrname}{Nimbus}
\newcommand{\cxo}{Chandra}
\newcommand{\xmm}{{\it XMM-Newton}}

\newcommand{\x}{4XMM\,J163159.8$-$47215}

\newcommand{\ef}{944\,MHz}
\newcommand{\tf}{1367\,MHz}
\newcommand{\mf}{1284\,MHz}
\newcommand{\al}{5500\,MHz}
\newcommand{\ah}{9000\,MHz}
\newcommand{\wf}{12\,$\mu$m}
\newcommand{\ebs}{$16.3\arcsec\times13.3\arcsec$}
\newcommand{\mbs}{$8\arcsec\times8\arcsec$}
\newcommand{\tbs}{$8.8\arcsec\times7.4\arcsec$}
\newcommand{\albs}{$6\arcsec\times6\arcsec$}
\newcommand{\ahbs}{$3\arcsec\times3\arcsec$}
\newcommand{\absmax}{$1.0\arcsec\times0.7\arcsec$}

\newcommand{\D}{$^{\circ} $}
\def\arcmin{\hbox{$^{\prime}$}}
\def\arcsec{\hbox{$^{\prime\prime}$}}

\usepackage{scalerel}
\newcommand\hi{H\,\protect\scaleto{$I$}{1.2ex}}
\newcommand{\hii}{H\,\protect\scaleto{$II$}{1.2ex}}
\newcommand{\ergcm}[1]{erg cm$^{-2}$ s$^{-1}$}
\newcommand{\ergs}[1]{$\times 10^{#1}$ erg s$^{-1}$}
\newcommand{\mjb}{\,mJy\,beam$^{-1}$}
\newcommand{\mujb}{\,$\mu$Jy\,beam$^{-1}$}

\begin{document}

\title{EMU discovery of \pwnname: a bow-shock PWN powered by \psr\ \\ escaping \snrname\ SNR~(\snr)}

\author[orcid=0000-0001-6109-8548]{S.\,Lazarevi\'c}
\affiliation{Western Sydney University, Locked Bag 1797, Penrith South DC, NSW 2751, Australia}
\affiliation{Australia Telescope National Facility, CSIRO, Space and Astronomy, PO Box 76, Epping, NSW 1710, Australia}
\affiliation{Astronomical Observatory, Volgina 7, 11060 Belgrade, Serbia}
\email[show]{s.lazarevic@westernsydney.edu.au}  

\author[0000-0002-0766-7313]{C.\,Maitra} 
\affiliation{Inter University Centre for Astronomy \& Astrophysics, Ganesh-khind, Pune 411007, India}
\affiliation{Max-Planck-Institut für extraterrestrische Physik, Gießenbachstraße 1, D-85748 Garching bei München, Germany}
\email[show]{chandreyee.maitra@iucaa.in}

\author[0000-0002-0457-3661]{A.\,Ahmad}
\affiliation{Western Sydney University, Locked Bag 1797, Penrith South DC, NSW 2751, Australia}
\email{}

\author[0000-0001-5609-7372]{R.\,Z.\,E.\,Alsaberi}
\affiliation{Western Sydney University, Locked Bag 1797, Penrith South DC, NSW 2751, Australia}
\affiliation{Faculty of Engineering, Gifu University, 1-1 Yanagido, Gifu 501-1193, Japan}
\email{}

\author[0000-0002-9618-2499]{S.\,Dai}
\affiliation{Australia Telescope National Facility, CSIRO, Space and Astronomy, PO Box 76, Epping, NSW 1710, Australia}
\email{}

\author[0000-0002-4814-958X]{D.\,Leahy}
\affiliation{Department of Physics and Astronomy, University of Calgary, Calgary, Alberta, T2N 1N4, Canada}
\email{} 

\author[0000-0002-4990-9288]{M.\,D.\,Filipovi\'c}
\affiliation{Western Sydney University, Locked Bag 1797, Penrith South DC, NSW 2751, Australia}
\email{}

\author[0000-0002-8186-4753]{P.\,G.\,Edwards}
\affiliation{Australia Telescope National Facility, CSIRO, Space and Astronomy, PO Box 76, Epping, NSW 1710, Australia}
\email{} 

\author[0000-0002-6447-4251]{O.\,Kargaltsev}
\affiliation{Department of Physics, The George Washington University, 725 21st St. NW, Washington, DC 20052, USA}
\email{}

\author[0000-0002-4441-7081]{M.\,Abdelmaguid}
\affiliation{Center for Astrophysics \& Space Science~(CASS), NYU Abu Dhabi, P.O. Box 129188, Abu Dhabi, UAE}
\email{} 

\author[0000-0003-4679-1058]{J.\,D.\,Gelfand}
\affiliation{Center for Astrophysics \& Space Science~(CASS), NYU Abu Dhabi, P.O. Box 129188, Abu Dhabi, UAE}
\email{} 

\author[0000-0001-7722-8458]{J.\,West}
\affiliation{School of Natural Sciences, University of Tasmania, PO Box 807, Sandy Bay, TAS 7006 Australia}
\affiliation{Dominion Radio Astrophysical Observatory, Herzberg Astronomy \& Astrophysics, National Research Council Canada, P.O. Box 248, Penticton, BC V2A 6J9, Canada}
\email{}

\author[0000-0002-6952-9688]{S.\,Hutschenreuter}
\affiliation{University of Vienna, Department of Astrophysics, Türkenschanzstraße 17, 1180, Vienna, Austria}
\email{} 

\author[0000-0002-8312-6930]{R.\,Brose}
\affiliation{Institute of Physics and Astronomy, University of Potsdam, 14476 Potsdam-Golm, Germany}
\email{}

\author[0000-0001-5953-0100]{R.\,Kothes}
\affiliation{Dominion Radio Astrophysical Observatory, Herzberg Astronomy \& Astrophysics, National Research Council Canada, P.O. Box 248, Penticton, BC V2A 6J9, Canada}
\email{}

\author[0000-0002-0416-3267]{V.\,Velovi\'c}
\affiliation{Western Sydney University, Locked Bag 1797, Penrith South DC, NSW 2751, Australia}
\email{}

\author[0000-0002-7239-2248]{C.\,Burger-Scheidlin}
\affiliation{Astronomy \& Astrophysics Section, School of Cosmic Physics, Dublin Institute for Advanced Studies, DIAS Dunsink Observatory, Dublin D15 XR2R, Ireland} 
\email{}

\author[0009-0003-2088-9433]{B.\,D.\,Ball}
\affiliation{Department of Physics, University of Alberta, Edmonton, Alberta, T6G 2E1, Canada}
\email{}

\author[0009-0003-0926-8791]{G.\,Graham}
\affiliation{Department of Physics, University of Alberta, Edmonton, Alberta, T6G 2E1, Canada}
\email{}

\author[0000-0002-6097-2747]{A.\,M.\,Hopkins}
\affiliation{School of Mathematical and Physical Sciences, Macquarie University, 12 Wally’s Walk, Macquarie Park, 2109, NSW, Australia}
\email{} 

\author[0000-0002-9516-1581]{G.\,P.\,Rowell}
\affiliation{School of Physics, Chemistry and Earth Sciences, The University of Adelaide, Adelaide, 5005, Australia}
\email{}

\author[0000-0001-9414-175X]{S.\,F.\,Rahman}
\affiliation{Syed Babar Ali School of Science and Engineering, Lahore University of Management Sciences, Lahore, Pakistan}
\email{}

\author[0009-0009-7061-0553]{Z.\,J.\,Smeaton}
\affiliation{Western Sydney University, Locked Bag 1797, Penrith South DC, NSW 2751, Australia}
\email{}

\author[0009-0001-6908-2433]{S.\,Taziaux}
\affiliation{Ruhr University Bochum, Faculty of Physics and Astronomy, Astronomical Institute~(AIRUB), Universitätsstraße 150, 44801 Bochum, Germany}
\affiliation{Australia Telescope National Facility, CSIRO, Space and Astronomy, PO Box 1130, Bentley WA 6102, Australia}
\email{}

\begin{abstract}

We report the discovery of a bow-shock pulsar wind nebula~(PWN), dubbed \pwnname, powered by the radio pulsar \psr\ and projected within the Galactic supernova remnant~(SNR) \snr~(\snrname). The system was first identified in observations from the Australian Square Kilometre Array Pathfinder~(ASKAP) Evolutionary Map of the Universe~(EMU) survey and further characterised using MeerKAT Galactic Plane Survey data together with follow-up observations at 5.5 and 9\,GHz obtained with the Australia Telescope Compact Array~(ATCA). Assuming a distance of 7\,kpc, the radio images resolve an elongated $\sim$80\arcsec~(2.7\,pc) cometary nebula, indicative of a high-velocity pulsar. An X-ray counterpart extending $\sim$50\arcsec~(1.7\,pc) is detected in archival \xmm\ data. 
The flat radio spectrum~($\alpha$\,=\,$-$0.27\,$\pm$\,0.05) and hard X-ray photon index~($\Gamma$\,=\,1.6\,$\pm$\,0.4) indicate synchrotron emission from relativistic particles injected in the pulsar wind. Polarisation analysis reveals a highly ordered magnetic field aligned with the nebular flow, with fractional polarisation reaching up to 30\% in the tail. An equipartition estimate gives a PWN magnetic-field strength of B$_{\rm eq}$\,$\approx$\,54$-$140\,$\mu$G. Pulsar timing over a $\sim$2.2\,yr baseline reveals strong timing noise and a small spin glitch with amplitude $\Delta\nu/\nu$\,=\,1.10\,$\times10^{-8}$. The SNR shows no clear diffuse X-ray counterpart. The morphology and multiwavelength properties of the \snrname--\pwnname\ system, together with evolutionary models, constrain the system age to $\sim$\,30--45\,kyr, placing the remnant in the late Sedov phase approaching the transition to the radiative stage.

\end{abstract}



\section{Introduction}
\label{intro}

\begin{figure*}
\centering
\includegraphics[trim=0 0 10 0, width=0.49\linewidth]{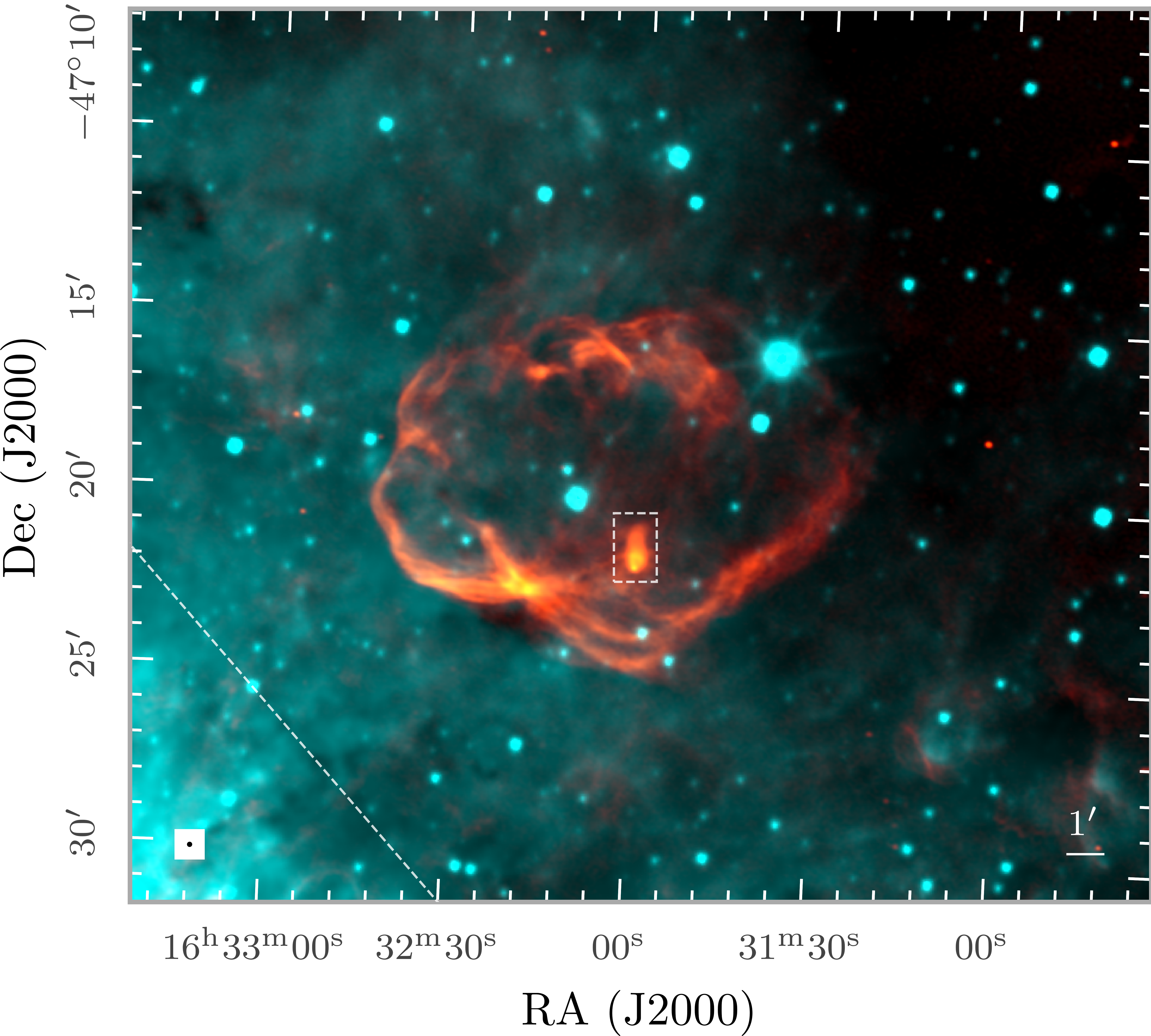}
\includegraphics[trim=10 0 0 0, width=0.49\linewidth]{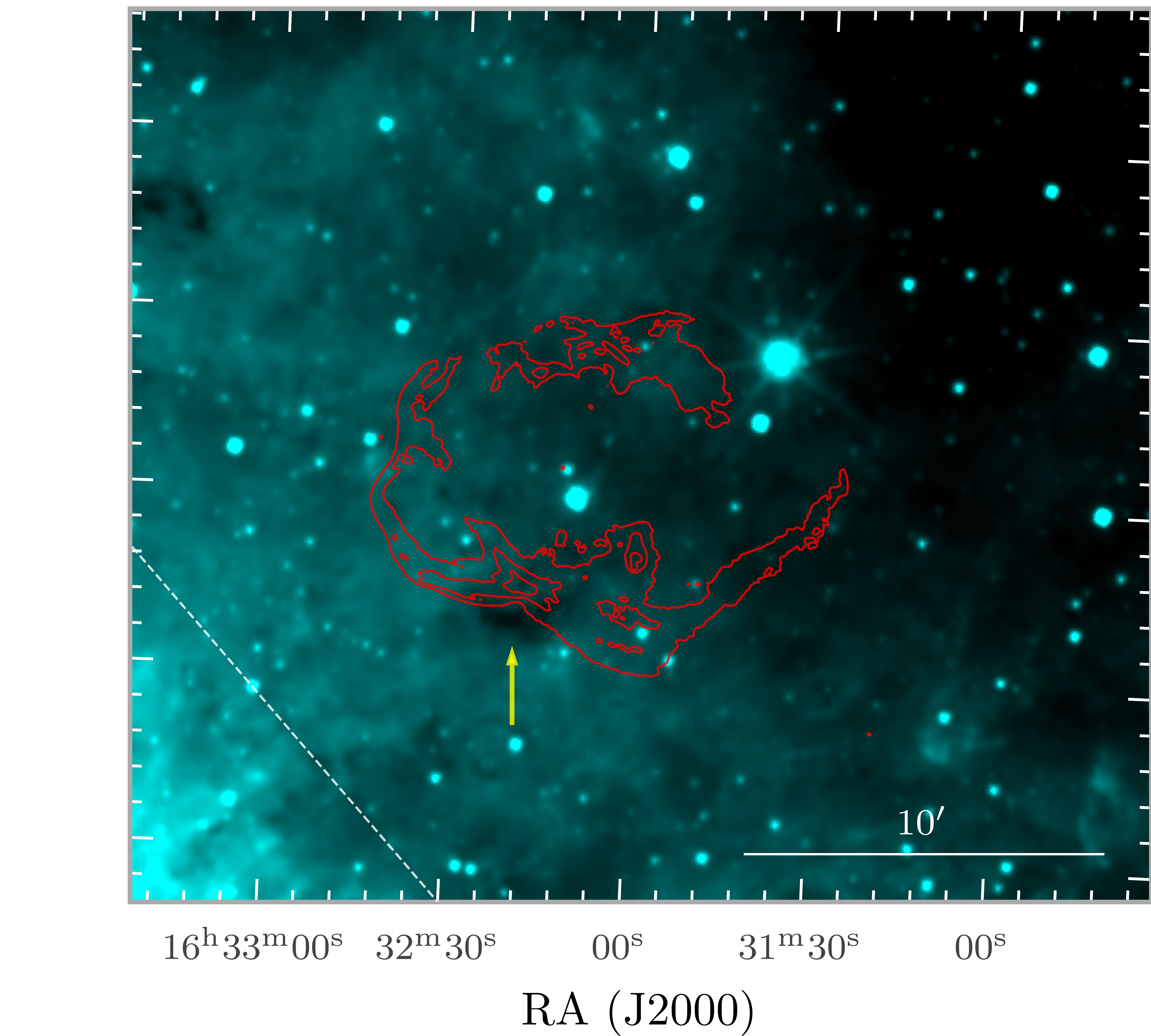}
\caption{Left: Composite image of ASKAP \tf\ radio emission~(orange-red) and WISE \wf\ infrared emission~(cyan) of the SNR\,\snr~(\snrname) region. PWN\,\pwn~(\pwnname) is marked by the white dashed box. The radio synthesised beam size of \tbs\ is shown in the lower-left corner. The dashed line indicates a line parallel to the Galactic plane, offset by 0.35\D. Right: WISE \wf\ map with overlaid radio contours at 1, 3, and 7\,\mjb. The yellow arrow points to the infrared dark region discussed in Section~\ref{sec:res-mir}.
}
\label{fig:fig1-rgb}
\end{figure*}

Massive stars end their lives in core-collapse supernova explosions, leaving behind expanding supernova remnants~(SNRs) and, in many cases, rapidly rotating neutron stars observed as pulsars. Pulsars lose a significant fraction of their rotational energy through highly relativistic, magnetised particle winds. The interaction of this outflow with the surrounding medium produces a pulsar wind nebula~\citep[PWN; see][for reviews]{2006ARA&A..44...17G,2017SSRv..207..175R}, observable across a broad frequency range from radio to $\gamma$-rays~\citep{book2}. 

The emission from PWN is primarily non-thermal. At radio and X-ray energy bands, the radiation is dominated by synchrotron emission from relativistic electrons and positrons spiralling in the nebular magnetic field. Higher-energies are typically produced via inverse Compton scattering, where the same particle population up-scatters ambient low-energy photons such as the cosmic microwave background, infrared dust emission, and stellar radiation fields~\citep[see][and references therein]{2017ASSL..446..161G}. 

PWNe trace both the energetics of the pulsar and the physical conditions of the surrounding medium~\citep{2017ASSL..446....1K}. In young SNR$-$PWN systems, the nebula is embedded within the freely expanding ejecta, and is generally well described as a quasi-spherical bubble of pulsar wind material centred on the neutron star~\citep{2008AIPC..983..171K}. As the SNR evolves, the reverse shock 
propagates back toward the SNR centre and eventually interacts with the PWN. Due to the natal kick imparted to the neutron star during the supernova explosion, combined with asymmetric reverse-shock propagation caused by density inhomogeneities in the surrounding interstellar medium~(ISM), different regions of the PWN may encounter the heated ejecta at different times. 
This interaction can significantly modify the nebular morphology and energetics. The PWN expansion slows in the direction of interaction, often producing an offset between the neutron star and the nebular centre~\citep{2022hxga.book...61M}. As the reverse shock engulfs the PWN, compression by the surrounding hot gas can amplify magnetic fields, increase synchrotron cooling rates, and produce substantial changes in the broadband spectral energy distribution~\citep[e.g.,][]{2020MNRAS.499.2051B,2023MNRAS.520.2451B,2026arXiv260515359D}. 

At later evolutionary stages, the pulsar can migrate toward or beyond the SNR shell. In this regime, the pulsar wind interacts directly with the ambient ISM and becomes confined by ram pressure, producing a bow-shock PWN characterised by a compact head and an elongated cometary tail aligned opposite to the pulsar’s direction of motion~\citep{2017JPlPh..83e6301K}. This morphology typically develops once the pulsar approaches or moves beyond the edge of the SNR, where the pulsar motion relative to the surrounding medium becomes supersonic~\citep[e.g.,][]{2002A&A...393..629B,2003A&A...397..913V}. Given that higher-energy particles cool more rapidly via synchrotron losses, X-ray-emitting electrons trace only the most recently accelerated population, while radio-emitting electrons can survive for significantly longer timescales. As a result, evolved PWNe may remain prominent at radio wavelengths even when X-ray emission becomes weak or undetectable, preserving a record of the integrated PWN history. 

Systems observed during the evolutionary transition between reverse-shock interaction and the formation of a bow-shock PWN provide valuable laboratories for studying the coupled evolution of neutron stars and their parent remnants. 
However, such associations are rare and observationally challenging to identify. As SNR shells fade with time, evolved systems can become faint, highly asymmetric, and easily confused with surrounding diffuse emission. 

The new generation of wide-area radio continuum surveys conducted with the Australian Square Kilometre Array Pathfinder~\citep[ASKAP;][]{2021PASA...38....9H} and MeerKAT~\citep{meerKAT} have renewed interest in known Galactic SNRs by enabling detailed studies of their morphology and internal structure. 
The combination of improved surface-brightness sensitivity and arcsecond-scale resolution is crucial for identifying and examining PWNe embedded within complex SNR or ISM environments. 

In this work, we report the discovery of a bow-shock PWN associated with the known Galactic SNR G336.7+0.5~\citep{1970AuJPA..14..133S}. The previously unresolved nebula exhibits a cometary morphology consistent with ram-pressure confinement and is powered by a young, energetic radio pulsar, \psr~\citep{2025MNRAS.537.2868A}. Hereafter, we refer to the SNR as \snrname\footnote{\snrname\ is a Latin word for a rain-bearing cloud and, in this context, reflects the extended, diffuse appearance of the SNR, reminiscent of a cloud-like structure. These clouds are commonly associated with stormy weather, naturally pairing with the name \pwnname\ given to the PWN.\label{nickname}} and the PWN as \pwnname\footref{nickname}. 

The paper is structured as follows. The observations and data processing procedures are described in Section~\ref{sec:data}. In Section~\ref{sec:results}, we present a multi-wavelength analysis of \pwnname–\snrname\ system, with a focus on the radio and X-ray properties of the PWN. The broader implications of these results are discussed in Section~\ref{sec:discussion}. We close by summarising our findings in Section~\ref{sec:conclusion}.

\section{Observations and data reduction}
\label{sec:data}

The PWN \pwnname\ was initially identified in ASKAP pilot survey observations~(Figure~\ref{fig:fig1-rgb}). Following its discovery, \citet{2025MNRAS.537.2868A} performed deep high time-resolution observations with Murriyang~(Parkes) radio telescope, leading to the detection of the powering pulsar, \psr. The pulsar is now part of a regular monitoring programme, and updated timing results are included in this work. We subsequently observed the system with the Australia Telescope Compact Array~(ATCA) and incorporated archival MeerKAT observations to characterise the radio emission across a broad frequency range. We also searched for high-energy counterparts and identified associated X-ray emission in \xmm. 
In addition, we used mid infrared~(MIR) data from the Wide-field Infrared Survey Explorer~\citep[WISE;][]{2010AJ....140.1868W} to investigate potential thermal emission. A summary of the observations used in this study is given in Table~\ref{tab:tab1-observationSummary}, with further details provided in the following sections. 

\begin{deluxetable}{l c c l}
\tablecaption{Summary of the observations used in this study.} 
\label{tab:tab1-observationSummary}
\tablehead{
\colhead{Telescope} & \colhead{Frequency} & \colhead{Resolution} & \colhead{Date} \\
\colhead{} & \colhead{\lbrack MHz\rbrack} & \colhead{\lbrack\arcsec\rbrack} & 
}
\startdata
ASKAP   & 944  & 16.3\,$\times$\,13.3                  & 2021\,Sep\,11 \\ 
MeerKAT & 1284 & 8\,$\times$\,8                        & 2018\,Aug\,26 \\
ASKAP   & 1368 & 8.8\,$\times$\,7.4                    & 2022\,Mar\,5  \\
ATCA    & 5500$^{\color{blue}a}$ & \multirow{2}{*}{6\,$\times$\,6$^{\color{blue}c}$} & 2024\,Jan\,12 \\
ATCA    & 5500$^{\color{blue}b}$ &                                & 2024\,Mar\,25 \\
ATCA    & 9000$^{\color{blue}a}$ & \multirow{2}{*}{6\,$\times$\,6$^{\color{blue}c}$} & 2024\,Jan\,12 \\
ATCA    & 9000$^{\color{blue}b}$ &                                & 2024\,Mar\,25 \\
Parkes  & 704$-$4032 & 468\,$\times$\,468$^{\color{blue}d}$       & 2023\,Nov\,8$-$\\
\xmm\,-pn   & 0.2$-$12$^{\color{blue}e}$ & 13\,$\times$\,13$^{\color{blue}f}$ & see Table~\ref{tab:tab2-x-ray} \\
\enddata
\tablecomments{{\color{blue}$^{a}$}ATCA observations obtained in the EW367 configuration. {\color{blue}$^{b}$}ATCA observations obtained in the 6A configuration. {\color{blue}$^{c}$}Resolution derived from the combined datasets. {\color{blue}$^{d}$}Resolution corresponds to the Murriyang multibeam receiver at 1400\,MHz. {\color{blue}$^{e}$}\xmm\ energy band is given in keV. {\color{blue} $^{f}$}On-axis angular resolution of the EPIC-pn at 1.5\,keV.}
\end{deluxetable}

\subsection{ASKAP}
\label{sec:askap}
We used ASKAP observations of \snrname\ obtained as part of a pilot programme of the Evolutionary Map of the Universe survey~\citep[EMU; project code AS101;][]{emu,norris20,2025PASA...42...71H} and a technical commissioning project~\citep[project code AS113;][]{test}. The observations were conducted in continuum mode at central frequencies of 944 and \tf, using the full instantaneous bandwidth of 288\,MHz. The corresponding scheduling blocks~(SBs) are SB32043 and SB37909, respectively. SB32043 was observed on 2021 September 11 using 34 of the 36 ASKAP antennas, while SB37909 was observed on 2022 March 5 with 35 antennas in operation. Each antenna has a diameter of 12\,m and is equipped with a phased array feed~\citep[PAF;][]{2012SPIE.8444E..2AS} mounted at the prime focus, providing a field of view~(FoV) of approximately 30\,deg$^{2}$~\citep[$\sim$6\D\,$\times$\,5\D;][]{2021PASA...38....9H}. The ASKAP baselines range from 22\,m to 6.4\,km. 

\begin{figure*}[!htp]
\includegraphics[trim=0 0 0 0, width=\textwidth]{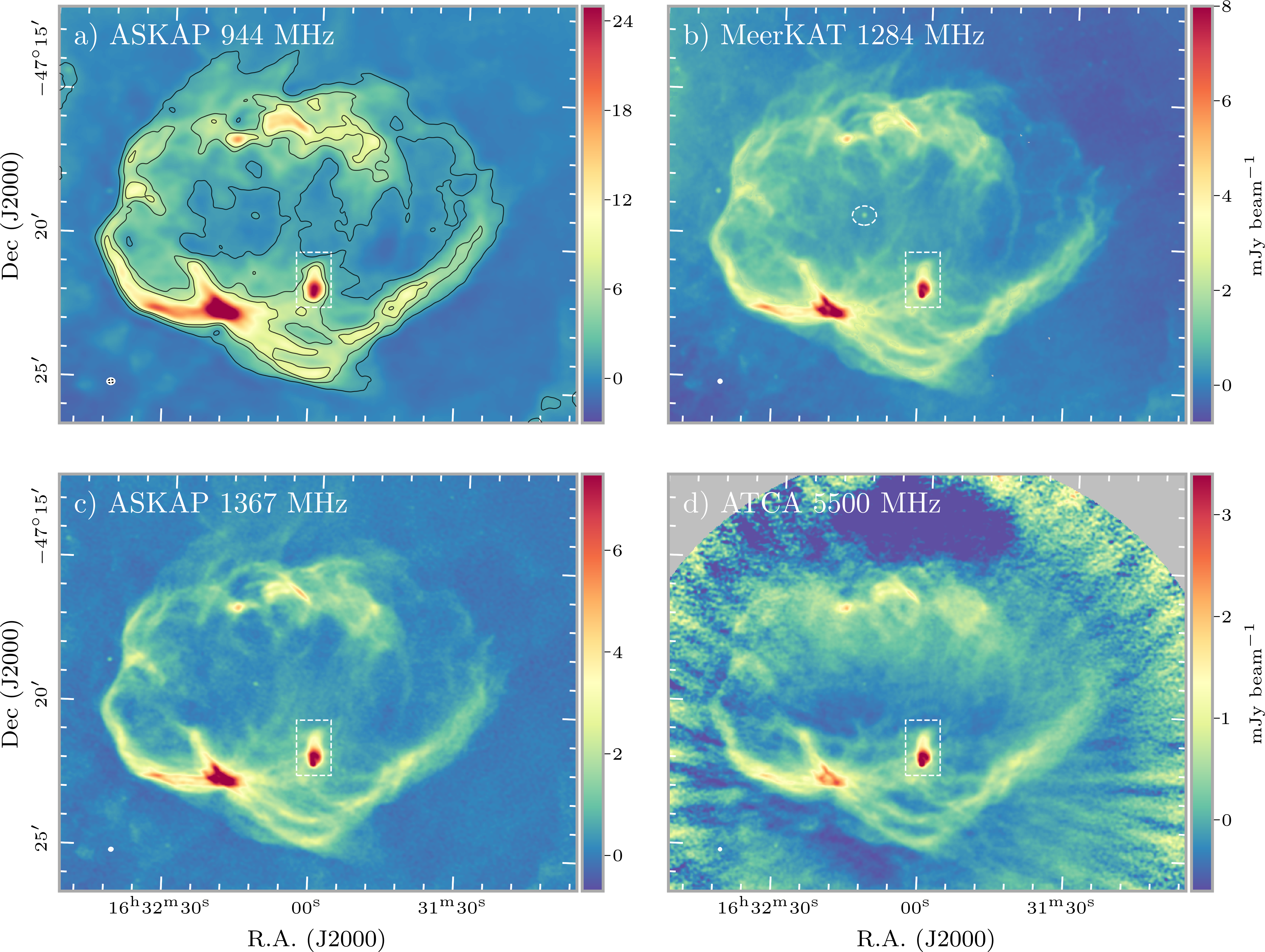}
\caption{Radio continuum images of \snrname\ SNR~(\snr) and the bow-shock PWN \pwnname~(\pwn), highlighted by the white dashed box. Panels show (a) ASKAP \ef~(resolution \ebs) with contours at 5, 20, and 37$\sigma$, where $\sigma = 2 \times 10^{-4}$\,Jy\,beam$^{-1}$, (b) MeerKAT \mf~(\mbs), (c) ASKAP \tf~(\tbs) and (d) ATCA \al~(\albs). The synthesised beam for each observation is shown as a white ellipse in the lower-left corner. The white ellipse in panel (b) marks the position of the OH\,336.739+0.522 maser~\citep[][]{2016ApJS..227...26Q}.
}
\label{fig:fig2-snrAllFreq}
\end{figure*}

The \ef\ image has a synthesised beam of \ebs\ with a position angle~(PA) of 84.5\D\ and a root-mean-square~(rms) noise level of 131\mujb~(Figure~\ref{fig:fig2-snrAllFreq}a). The \tf\ image has a higher angular resolution of \tbs\ with a PA of 78.7\D\ and a noise level of 68\mujb~(Figure~\ref{fig:fig2-snrAllFreq}b). Both datasets have a total integration time of 10\,h. Data calibration and imaging were performed using the ASKAPsoft data processing pipeline~\citep{Guzman_Askapsoft} at the Pawsey Supercomputing Centre. Due to the shortest baselines of ASKAP, emission on angular scales larger than approximately 43\arcmin$-$\,60\arcmin\ is resolved out in the EMU images~\citep{2025PASA...42...71H}. The science-ready data products are publicly available through the CSIRO ASKAP Data Science Archive~(CASDA)\footnote{\href{https://data.csiro.au/domain/casdaObservation}{CSIRO ASKAP Data Science Archive~(CASDA)}}. 

\subsection{MeerKAT}
\label{sec:meerkat}

The MeerKAT observations of \snrname\ 
were obtained from the SARAO MeerKAT Galactic Plane Survey~\citep[SMGPS;][]{2024MNRAS.531..649G,2025A&A...695A.144B}. The MeerKAT array consists of 64 antennas, each with a dish diameter of 13.5\,m, and baselines ranging from 29\,m to 8\,km. SMGPS is a continuum L-band survey covering frequencies from 886$-$1678\,MHz, centred at \mf. The survey is organised into observing tiles, each covering a sky area of 3\D\,$\times$\,3\D. A detailed description of the SMGPS observing strategy and data reduction procedures is given by \citet{2024MNRAS.531..649G}. 

A SMGPS tile containing \snrname\ region was observed on 2018 August 26, with 61 antennas contributing to the observation. The final total intensity image has the survey’s common angular resolution of \mbs, 
and reaches a local rms noise level of 56\mujb~(Figure~\ref{fig:fig2-snrAllFreq}b). 

Due to the lack of zero-spacing information, the sensitivity of MeerKAT to extended emission is limited to angular scales of approximately 21\arcmin$-$\,40\arcmin\ across the L-band, meaning that very large-scale structures may suffer from missing flux. 
While MeerKAT’s longer baselines provide higher angular resolution than ASKAP, its sensitivity to emission on the largest angular scales is lower. The ASKAP EMU observations retain better sensitivity to large-scale diffuse emission, making the two facilities complementary. All SMGPS DR1 data products are publicly available~\citep{meerkatData}.

\subsection{ATCA}
\label{sec:atca}

We conducted follow-up continuum observations of \pwnname\ PWN with ATCA, under project code C3577~(PI: S.~Lazarevi\'c). The ATCA array consists of six 22-m antennas. Observations were performed at central frequencies of 5500 and \ah, each with a bandwidth of 2048\,MHz, using the Compact Array Broadband Backend~(CABB). Data were obtained in two array configurations\footnote{\href{https://www.narrabri.atnf.csiro.au/operations/array\_configurations/configurations.html}{ATCA array configurations}}: EW367 on 2024 January 12 and 6A on 2024 March 25. The shortest baseline is set by the EW367 configuration~($\approx$\,46\,m), which limits sensitivity to structures larger than 4\arcmin$-$\,6\arcmin. The observations totalled $\sim$12.5~h of integration time in each band. The target field was observed as a single pointing centred on \pwnname\ PWN, with full Stokes products~($I, Q, U, V$) recorded. 

We used the standard southern sky calibrator PKS\,B1934$-$638 to set the absolute flux density scale and to correct for the frequency-dependent instrumental bandpass. The secondary calibrator, PKS\,1101$-$536, located 45\arcmin\ from the pointing centre, was observed every 25\,min to track time-dependent complex gain and phase variations. Details of the calibrators are available through the ATCA calibrator database\footnote{\href{https://www.narrabri.atnf.csiro.au/calibrators/calibrator_database.html}{ATCA calibrator database}}. 

Data reduction and imaging were performed using standard procedures within the \texttt{MIRIAD}\footnote{\href{http://www.atnf.csiro.au/computing/software/miriad/}{\texttt{MIRIAD} data reduction package}} data reduction package~\citep{1994AAS..108..585S}. 
Initial inspection and flagging were carried out to remove radio-frequency interference~(RFI) and corrupted visibilities. The calibrated datasets from the two array configurations were then combined to improve~($u,v$)-coverage and overall sensitivity. Imaging was completed using the \texttt{INVERT} task with multi-frequency synthesis and Briggs weighting, adopting a robust parameter of $-$0.5. This value balances angular resolution and surface-brightness sensitivity while suppressing sidelobe structure. 
Deconvolution was performed using \texttt{MFCLEAN} separately for each Stokes parameter. The clean components were then restored with \texttt{RESTOR}, applying a common Gaussian restoring beam of 6\arcsec$\times$\,6\arcsec\ and position angle of 0\D, ensuring consistent angular resolution across all Stokes images. Primary-beam correction was applied using \texttt{LINMOS}. 
The final \snrname\ Stokes~$I$ \al\ images is shown in Figure~\ref{fig:fig2-snrAllFreq}d. 

\begin{figure*}
\includegraphics[trim=0 0 0 0, width=\textwidth]{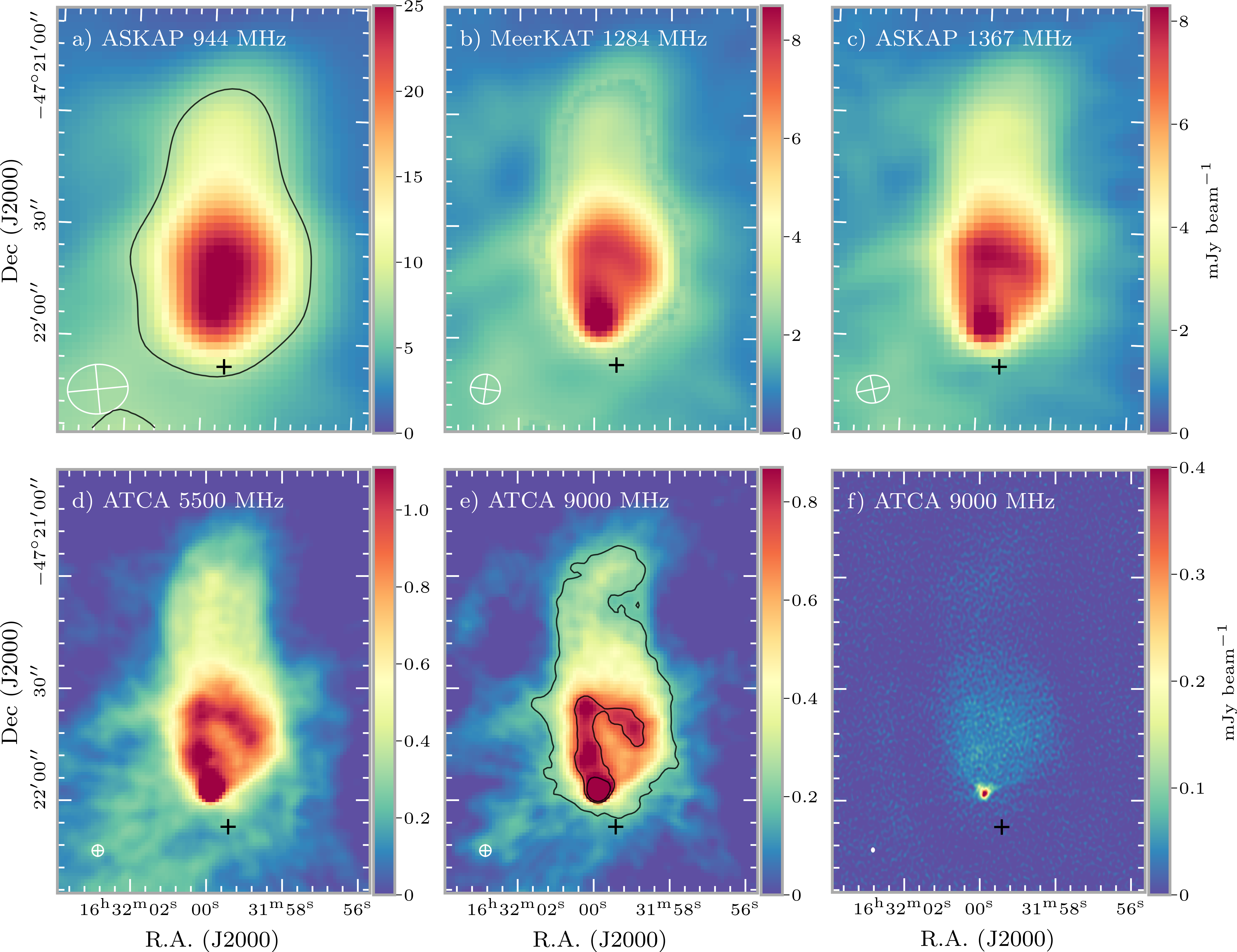}
\caption{Radio continuum images of \pwnname\ PWN. Panels show (a) ASKAP \ef~(\ebs\ resolution) with a contour at 7.4\,mJy\,beam$^{-1}$, (b) MeerKAT \mf~(\mbs), (c) ASKAP \tf~(\tbs), (d) ATCA \al~(\ahbs), (e) ATCA \ah~(\ahbs) with contours at 0.2, 0.75 and 1\,mJy\,beam$^{-1}$ and (f) ATCA \ah~(\absmax). All images were produced using Briggs weighting with robust parameter of $-$0.5, except panel (f) where robust of $-$2 was used. The synthesised beam for each observation is shown as a white ellipse in the lower-left corner. The position of \psr, as inferred from pulsar timing, is indicated by the black cross. The apparent offset between the timing position and the radio peak is discussed in Section~\ref{sec:res-pulsar}.}
\label{fig:fig3-pwnAllFreq}
\end{figure*}

For quantitative analysis, we used 6\arcsec$\times$\,6\arcsec\ images, while for morphological analysis, we produced additional higher-resolution images. The 3\arcsec$\times$\,3\arcsec\ images were generated using the same robust weighting to better resolve the internal nebular structure~(Figures~\ref{fig:fig3-pwnAllFreq}d~and~\ref{fig:fig3-pwnAllFreq}e), and a \absmax\ was produced using uniform weighting~(a robust parameter of $-2$) to probe the highest-resolution morphology of the PWN point-like source and its immediate environment~(Figure~\ref{fig:fig3-pwnAllFreq}f). 
The observations are publicly available through the Australia Telescope Online Archive\footnote{\href{https://data.csiro.au/domain/atoaObservation}{Australia Telescope Online Archive~(ATOA)}}~(ATOA). 

\subsection{Murriyang/Parkes radio telescope}

An ongoing timing campaign of PSR J1631$-$4722 is being conducted with Murriyang, the CSIRO Parkes 64-m radio telescope. Observations are carried out in the pulsar fold mode using the Ultra-Wideband Low~(UWL; 704$-$4032\,MHz) receiver~\citep{hobs2020}. Full Stokes polarisation data are recorded with a frequency resolution of 1\,MHz and 1024 phase bins. 
Polarisation and absolute flux-density calibration are performed using a pulsed noise signal injected into the signal path prior to each observation. 

We manually excised data affected by narrow-band and impulsive radio-frequency interference (RFI) for each sub-integration using \texttt{PSRCHIVE}~\citep{2004PASA...21..302H}. Polarisation and flux density calibration were performed following standard procedures within \texttt{PSRCHIVE}~\citep[e.g.,][]{2015MNRAS.449.3223D}. 
Each observation was averaged in time and frequency to form integrated pulse profiles, and pulse times of arrivals~(ToAs) were measured for each observation 
using \texttt{PSRCHIVE}. Timing analysis was carried out using the \texttt{TEMPO2} software package~\citep{2006MNRAS.369..655H}, see Section~\ref{sec:res-pulsar}.

\subsection{\xmm}
\label{sec:xmm} 

\begin{deluxetable}{c c c c c}
\tablecaption{\xmm\ observations of \pwnname\ PWN.}
\label{tab:tab2-x-ray}
\tablehead{
\colhead{Date} & \colhead{ObsID} & \colhead{Net Exposure \lbrack ks\rbrack} & \colhead{CCD Readout mode} & \colhead{Off-axis angle \lbrack\arcmin\rbrack} \\
\colhead{} & \colhead{} & \colhead{pn\,/\,MOS1\,/\,MOS2} & \colhead{pn\,/\,MOS1\,/\,MOS2} & \colhead{pn\,/\,MOS1\,/\,MOS2} 
}
\startdata
2011 Feb 2  & 0654190201 & 17.0 / 21.6 / 21.6 & Full Frame / Full Frame / Small Window & 6.2 / 6.2 / 6.1 \\ 
2014 Aug 24 & 0728560201 & 33.2 / 34.8 / 34.7 & Full Frame / Full Frame / Small Window & 5.8 / 6.8 / 7.5 \\ 
2014 Aug 26 & 0728560301 & 20.0 / 21.6 / 21.6 & Full Frame / Full Frame / Small Window & 5.7 / 6.6 / 7.3 \\ 
2018 Sep 18 & 0823990901 & 23.4 / $-$ / $-$     & Full Frame / Small Window / Timing     & 5.7 / $-$  / 7.2 \\ 
\enddata
\end{deluxetable}

Following the discovery of \pwnname, we searched X-ray archives for observations covering its sky position. Although no dedicated X-ray observations targeting the PWN exist, the source lies within the FoV of four \xmm\ observations~(ObsIDs 0654190201, 0728560201, 0728560301, and 0823990901), originally targeting the high-mass X-ray binary IGR J16328$-$4726~\citep{2013ApJ...762...19F}. Details of these observations are listed in Table~\ref{tab:tab2-x-ray}. 

Data from the three European Photon Imaging Camera detectors~\citep[EPIC;][]{2001A&A...365L..18S,2001A&A...365L..27T}: MOS1, MOS2, and pn, were processed using the \xmm\ data analysis software SAS version\footnote{\href{https://www.cosmos.esa.int/web/xmm-newton/sas-release-notes-2210}{Science Analysis Software~(SAS)}} 22.1.0. 
Periods of high background flaring activity were identified by extracting light curves in the 7.0$-$15.0\,keV energy range. Time intervals with background rates $\geq$\,8 and 2.5\,cts\,ks$^{-1}$\,arcmin$^{-2}$ for EPIC-pn and EPIC-MOS, respectively, were removed following the procedure of~\citet{2013A&A...558A...3S}. Event lists were filtered using the SAS task \texttt{evselect} with standard criteria~(\texttt{\#XMMEA\_EP \&\& PATTERN<=4} for EPIC-pn and \texttt{\#XMMEA\_EM \&\& PATTERN<=12} for EPIC-MOS). 

We extracted the spectrum of the hard X-ray source using a circular region with a radius of 30\arcsec\ centred on the source position. 
The background spectrum was taken from nearby circular regions with radii of 50\arcsec, selected away from the source and avoiding astrophysical sources or detector artefacts.

The SAS tasks \texttt{rmfgen} and \texttt{arfgen} were used to create the redistribution matrices and ancillary files for spectral analysis. The spectrum was grouped to ensure a minimum of one count per spectral bin. The spectral fitting was performed using the \texttt{XSPEC} package~\citep[version~12.14.1;][]{1996ASPC..101...17A} with the C-statistic. Errors were estimated at the 90\% confidence level.

No X-ray emission associated with \snrname\ was detected in the \xmm\ data. We also searched for the SNR diffuse emission in the SRG/eROSITA first all sky survey data~\citep[eRASS1;][]{2024AA...682A..34M}, but no corresponding X-ray counterpart was found. 
The observed count-rate upper limits are 0.018\,counts\,s$^{-1}$ from \xmm\ and 0.03\,counts\,s$^{-1}$ from eRASS1, the first all-sky survey data from SRG/eROSITA.

\begin{figure}
\centering
\includegraphics[width=0.75\columnwidth]{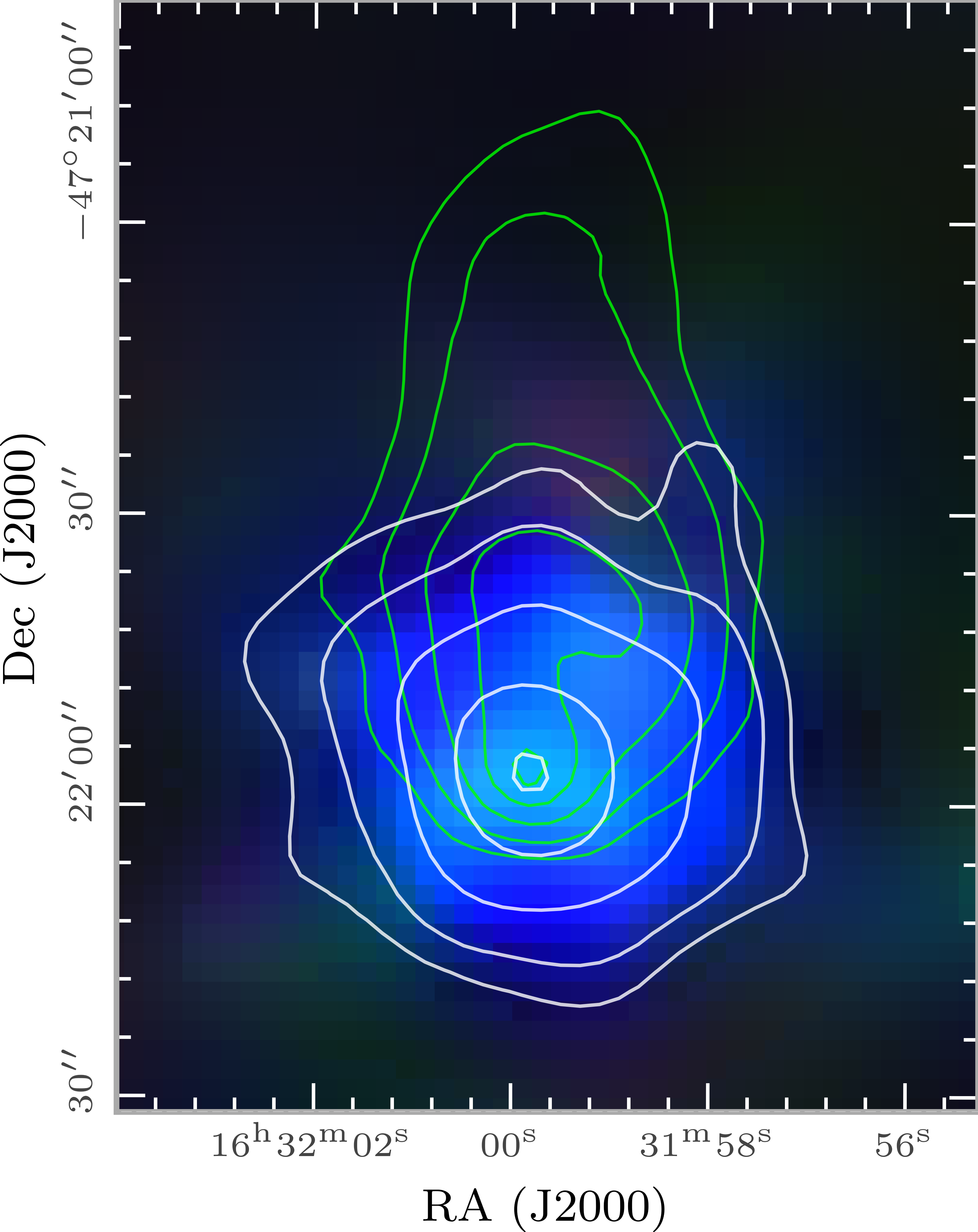}
\caption{\xmm\ X-ray RGB image of \pwnname, combining all observations listed in Table~\ref{tab:tab2-x-ray}. The image is constructed from the energy bands: 0.2$-$1 keV~(red), 1$-$2~keV~(green), and 2$-$4.5 keV~(blue). X-ray contours derived from the mosaic are shown in white, while radio contours tracing the PWN \tf\ emission are overlaid in green.
}
\label{fig:fig8-x-ray}
\end{figure}

\section{Results}
\label{sec:results}

As part of a systematic re-examination of known Galactic SNRs using ASKAP survey data, we identified a new PWN J1631$-$4721~(\pwnname), projected within SNR \snr~(\snrname). \pwnname$-$\snrname\ system is shown in Figures~\ref{fig:fig1-rgb} and \ref{fig:fig2-snrAllFreq}, with the PWN highlighted by the white dashed box. Follow-up pulsar searches 
with Murriyang, the CSIRO’s Parkes radio telescope, 
confirmed the presence of a young and energetic radio pulsar \psr~\citep{2025MNRAS.537.2868A}. The pulsar has a spin period of $P$\,=\,118.7\,ms and a period derivative of $\dot{P} = 5.6 \times 10^{-14}$, implying a characteristic age of $\tau_c \approx 34$ kyr and a spin-down luminosity of $\dot{E} = 1.3 \times 10^{36}\,\mathrm{erg\,s^{-1}}$. These properties indicate that \psr\ is sufficiently energetic to power the observed PWN. The confined morphology of the nebula suggests that \pwnname\ belongs to the small but growing class of bow-shock PWNe produced by supersonic pulsars moving through the ISM~\citep{2003A&A...397..913V}. 


\subsection{Radio morphology of \snrname\ and \pwnname}
\label{sec:res-morphology}

The radio morphology of \snrname\ SNR is broadly consistent with that reported at the time of its discovery~\citep{1970AuJPA..14..133S}. \pwnname\ PWN was not resolved in earlier radio continuum surveys and could not be distinguished from bright SNR structures due to the limited angular resolution of surveys such as MOST~\citep{1996A&AS..118..329W} and PMN~\citep{2007Ap&SS.307..423S}, see Section~\ref{sec:res-flux}. Here we examine the radio morphology of \snrname\ and \pwnname\ across multiple frequencies using new high-resolution radio observations~(Figures~\ref{fig:fig2-snrAllFreq}~and~\ref{fig:fig3-pwnAllFreq}). 

\snrname\ exhibits a roughly elliptical, shell 
morphology elongated along the east-west axis~(Figure~\ref{fig:fig1-rgb}). The shell is brightest and best defined along the southern and southeastern rims, where prominent filaments and bright clump-like structures are observed. In contrast, the northern and western rims are fainter and fragmented. This asymmetric brightness distribution likely reflects interaction with an inhomogeneous ISM, with an increasing ambient density toward the southeastern side, closer to the Galactic plane~(Figure~\ref{fig:fig1-rgb}). The flattened morphology parallel to the Galactic plane is consistent with a blast wave expanding into a strong ambient density gradient. Similar behaviour has been reported in other Galactic plane remnants~\citep[e.g., Diprotodon SNR;][]{2024PASA...41..112F}. Expansion into a lower-density environment away from the plane leads to more rapid evolution, producing a partial breakout, fainter shell. The interior emission is weak and diffuse. Overall, the radio morphology is consistent with an evolved shell-type SNR~(further discussed in Section~\ref{sec:dis-evolutionaryStatus}). 

We measured an angular diameter of approximately 12.0\arcmin$\times$\,14.4\arcmin\ for \snrname, with a centroid at (R.A.,\,Dec)$_\text{J2000}$ = 16:32:03.2, $-$47:19:02. This is slightly larger than the previously reported~\citep[$\sim$10\arcmin$\times$\,14\arcmin;][]{2025JApA...46...14G}, owing to the detection of faint northern filamentary emission not seen in earlier surveys. 
For an adopted distance of 7\,kpc~(see Section~\ref{sec:dis-distance}), this corresponds to a physical size of $\sim$24.5\,pc\,$\times$\,29.3\,pc. Very faint outflow-like emission extending northwest through the opening in the shell and directed away from the Galactic plane is detected in the ASKAP and MeerKAT images~(Figure~\ref{fig:fig2-snrAllFreq}). As its surface brightness is close to the noise level, this feature is not included in the size estimate. 

\pwnname\ is projected offset from the geometric centre of \snrname. Its morphology consists of a bright, compact head followed by diffuse emission that remains prominent before gradually fades, extending $\sim$80\arcsec\ northward. This angular extent corresponds to a projected physical length of 2.7\,pc at a distance of 7\,kpc~(see Section~\ref{sec:dis-distance}). The cometary appearance is characteristic of bow-shock PWNe, formed when a high-velocity pulsar moves supersonically through the ambient medium, producing a ram pressure confined nebula 
channelled opposite to the direction of motion. 
Similar structures are observed in other systems, such as the Mouse~\citep{2005AdSpR..35.1129Y}, 
the Cannonball PWN~\citep{2019ApJ...876L..17S} and Potoroo~\citep{2024PASA...41...32L}. 

A closer view of \pwnname~(Figure~\ref{fig:fig3-pwnAllFreq}) shows a conical structure dominated by a bright head-like feature at its apex, located at~(R.A.,\,Dec)$_\text{\rm J2000}$ = 16:31:59.8, $-$47:21:56. This compact emission is naturally interpreted as the radio counterpart to \psr, but the pulsar timing position lies ahead of the nebula, with a lateral offset of 12\arcsec. Although upstream offsets have been observed in other bow-shock PWNe~\citep[e.g.,][]{2010ApJ...712..596N,2012ApJ...746..105N}, 
the offset seen in \pwnname\ is most plausibly attributable to the pulsar positional uncertainty, given the high level of timing noise~(see Section~\ref{sec:res-pulsar}). 
This interpretation is further supported by the spatial coincidence between the radio and the X-ray emission peak~(Figure~\ref{fig:fig8-x-ray}). 

The ATCA 5500 and \ah\ images reveal at least four 
bright knots along the eastern edge and base of the conical structure. To better resolve these fine features, we reprocessed the ATCA data using a Gaussian restoring beam of 3\arcsec$\times$\,3\arcsec, as shown in  Figures~\ref{fig:fig3-pwnAllFreq}d~and~\ref{fig:fig3-pwnAllFreq}e. 
We also produced an additional \ah\ image~(Figure~\ref{fig:fig3-pwnAllFreq}f), using a robust parameter of $-$2 to maximise angular resolution and examine the nature of the apex emission. If the source were unresolved, the emission would appear point-like and a bow-shock interpretation would be difficult to reconcile with the observations. However, the observed extended structure supports a pulsar moving supersonically through the ambient medium.

\subsection{Radio--MIR correlation of Nimbus}
\label{sec:res-mir}

Comparison of radio and MIR emission is a commonly used technique for distinguishing non-thermal SNR emission from thermally emitting sources such as \hii\ regions~\citep[e.g.,][]{1996A&AS..118..329W,2021A&A...651A..86D,2023MNRAS.524.1396B,2025ApJ...988...75B,2025A&A...693A.247A}. While \hii\ regions exhibit strong MIR emission from warm dust and polycyclic aromatic hydrocarbons~(PAHs), supernova shocks can both transiently heat and efficiently destroy dust grains and large molecules, resulting in weak or absent MIR emission~\citep{2015ApJ...803....7S}. Using WISE 12\,$\mu$m data as a tracer of hot dust and PAHs, we find no MIR counterpart associated with either the SNR shell or the PWN~(Figure~\ref{fig:fig1-rgb}). 

A local MIR emission deficit is observed 
adjacent to the brightest part of the shell, indicated by the yellow arrow in Figure~\ref{fig:fig1-rgb}\,(right). 
This feature is most likely caused by a foreground or embedded dark cloud in which cold gas and dust absorb the bright Galactic plane MIR background. If physically interacting with the remnant, such a structure could also contribute to the enhanced radio brightness observed in this region. 

We also identify a compact radio source located east of the centre of \snrname, marked with a white ellipse in Figure~\ref{fig:fig2-snrAllFreq}b. This source has a clear infrared counterpart catalogued as an OH maser~\citep[OH\,336.739+0.522;][]{2016ApJS..227...26Q}. The detected emission corresponds to the 1612\,MHz OH satellite transition, which is typically associated with circumstellar envelopes of evolved giant stars rather than with SNR–molecular cloud interactions, which are commonly traced by the 1720\,MHz OH transition. We therefore treat this source as unrelated to the SNR. 

\subsection{PSR J1631--4722 spin and astrometric parameters}
\label{sec:res-pulsar}

%

\begin{figure}[!ht]
\centering
\includegraphics[trim=10 5 0 0,width=\columnwidth]{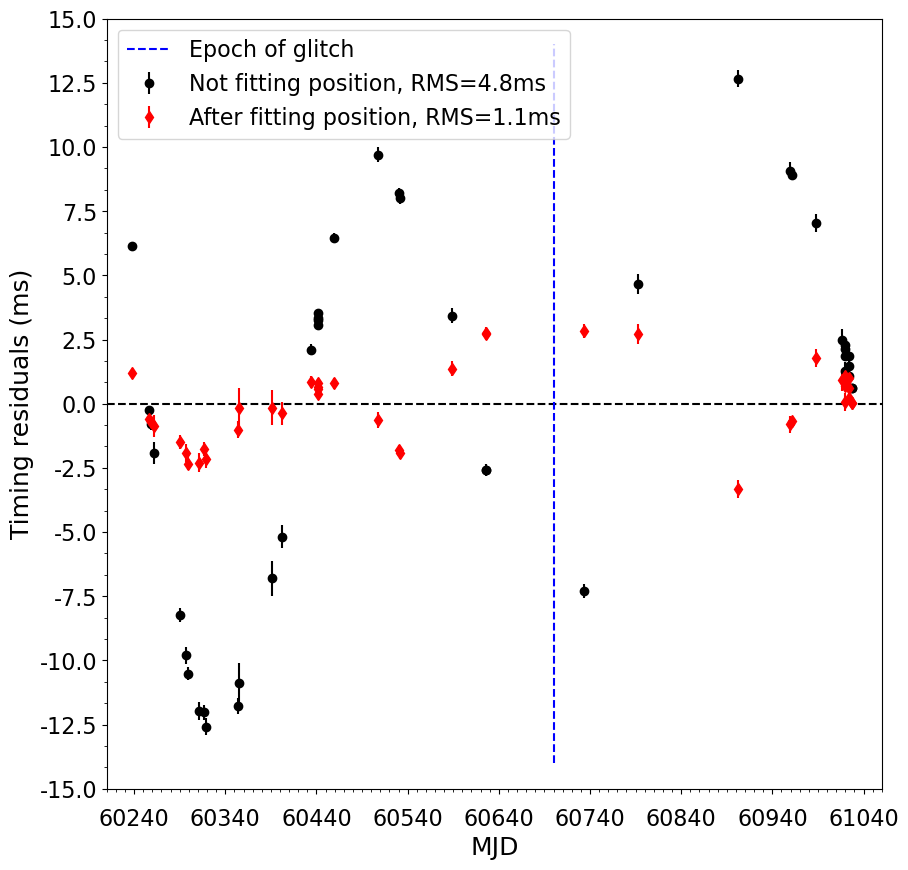}
\caption{Timing residuals of \psr. Black points with error bars show the timing residuals obtained without fitting for the pulsar position, while red points show the residuals after the fit. The blue dashed line marks the epoch of the glitch.}
\label{fig:psr-resi}
\end{figure}

We currently have a timing baseline of approximately 810~days~(2.2~yr) for \psr. In Figure~\ref{fig:psr-resi}, we show the timing residuals of \psr~(black points) obtained with the pulsar position fixed to that of the X-ray counterpart~(Section~\ref{sec:res-x-ray}). Clear variations on timescales of about one year are evident, indicating an offset in the assumed position.
After including the pulsar position as a free parameter in the fit, we obtain a timing position of (R.A.,\,Dec)$_\text{J2000}$ = 16:31:59.32(2), $-$47:22:03.4(4). The updated timing solution, including the fitted astrometric and spin parameters, is summarised in~(Table~\ref{tab:psr}). 
The resulting residuals are shown as red points. Although the rms of the residuals improves significantly after fitting for position, we caution that the derived timing position may be affected by unmodelled timing noise. For young pulsars such as \psr, strong timing noise on timescales from months to years is often observed~\citep[e.g.,][]{2019MNRAS.489.3810P}, which can make it difficult to obtain a reliable timing position with a relatively short observing baseline. Therefore, the actual positional uncertainty may be larger than indicated by the formal uncertainties reported in Table~\ref{tab:psr}. 

\begin{table}
\begin{center}
\caption{Measured and derived parameters of PSR J163--4722 from the current 810-day timing baseline. The reported values are an update of~\cite{2025MNRAS.537.2868A}, their Table\,1.}
\label{tab:psr}
\begin{tabular}{lcl}
\hline
\hline
\multicolumn{3}{c}{Measured parameters}     \\
\hline
R.A.$_\text{J2000}$       && 16:31:59.32(2) \\
Dec$_\text{J2000}$        && $-$47:22:03.4(4) \\
$\nu$ [Hz]                && 8.4232332246(4) \\
$\dot{\nu}$ [Hz\,s$^{-1}$]&& $-3.94381(1)\times10^{-12}$ \\
PEPOCH [MJD]              && 60237.254487 \\
Time span [MJD]           && 60208.046$-$61040.027 \\
DM [cm$^{-3}$\,pc]        && 873.75(7) \\
\hline
\multicolumn{3}{c}{Derived parameters}\\
\hline
$P$ [ms]                  && $118.719258192(6)$ \\
$\dot{P}$ [s\,s$^{-1}$]   && $5.55851(2)\times10^{-14}$ \\
$\tau_{\rm c}$ [kyr]      && $33.9$ \\
$\dot{E}$ [erg\,s$^{-1}$] && $1.3\times10^{36}$ \\
$B_{\rm s}$ [G]           && $2.6\times10^{12}$  \\
\hline
\end{tabular}
\end{center}
\end{table}

In addition to the timing noise, we detected a glitch in the pulsar spin period with an amplitude of $\Delta\nu/\nu$\,=\,1.10(2)$\times10^{-8}$ occurring between MJDs 60626 and 60733. Glitches of comparable amplitude are common among young pulsars~\citep{2021MNRAS.508.3251L}. Given the significant timing noise, the detected glitch, and the likelihood of future glitches, a longer timing baseline will be required before a more robust astrometric position can be obtained for \psr~\citep{2019MNRAS.489.3810P}. 

\subsection{Flux densities of \snrname\ and \pwnname}
\label{sec:res-flux}

In Table~\ref{tab:tab3-flux} we summarise the integrated flux densities of \snrname\ and \pwnname\ derived from the new radio observations presented in this work, together with values reported in previous studies. Using the ASKAP 944 and \tf\ and MeerKAT \mf\ data, we measure flux densities for the full SNR, the SNR shell, and the PWN. Due to the limited FoV of the ATCA 5500 and \ah\ observations relative to the angular extent of the remnant, reliable flux density measurements are possible only for the PWN. 

To ensure consistency across frequencies, all images were regridded to the finest pixel scale, that of the MeerKAT data~(1.5\arcsec$\times$\,1.5\arcsec), and then convolved to the largest synthesised beam, corresponding to the ASKAP \ef\ image~(\ebs). Integrated flux densities were extracted from identical regions defined by the 5$\sigma$ contour for \snrname\ and the 37$\sigma$ contour for \pwnname, as shown in Figure~\ref{fig:fig2-snrAllFreq}a. The flux density of the SNR shell was calculated as the difference between these two flux values. A conservative systematic uncertainty of 10\% on the absolute flux density scale is adopted throughout this work and included in the quoted uncertainties. 

Radio interferometers lack true zero-spacing information, and smooth low surface brightness emission can be partially resolved out, depending on the shortest baselines, imaging strategy, and the intrinsic angular scale of the source. As a result, integrated flux densities for extended structures may represent lower limits. In our case, this effect is expected to be modest, given that the average angular diameter of the remnant~($\approx$13.2\arcmin) lies well within the maximum recoverable angular scales of the ASKAP and MeerKAT observations~(see Section~\ref{sec:data}). The ASKAP and MeerKAT flux densities of \pwnname\ are expected to be unaffected, while missing flux effects in the ATCA measurements are likely negligible.

\begin{deluxetable}{c c c c}
\tablecaption{Integrated flux densities of \snrname$-$\pwnname\ system measured from ASKAP, MeerKAT and ATCA data and compiled from the MOST and PMN SNR catalogues. Dashes indicate measurements that could not be obtained because of the limited FoV or insufficient angular resolution to distinguish the PWN.}
\label{tab:tab3-flux}
\tablehead{
\colhead{Frequency} & \colhead{\snrname} & \colhead{\snrname\ shell} & \colhead{\pwnname} \\
\colhead{$\nu$ \lbrack MHz\rbrack} & \colhead{$S_{\rm SNR}$ \lbrack Jy\rbrack} & \colhead{$S_{\rm shell}$ \lbrack Jy\rbrack} & \colhead{$S_{\rm PWN}$ \lbrack Jy\rbrack}
}
\startdata
 843 & 6.1             & $-$             & $-$               \\ 
 944 & 5.8\,$\pm$\,0.6 & 5.6\,$\pm$\,0.6 & 0.16\,$\pm$\,0.02 \\ 
1284 & 5.3\,$\pm$\,0.5 & 5.2\,$\pm$\,0.5 & 0.15\,$\pm$\,0.02 \\
1368 & 4.5\,$\pm$\,0.5 & 4.3\,$\pm$\,0.4 & 0.15\,$\pm$\,0.01 \\
4850 & 0.6             & $-$             & $-$               \\
5500 & $-$             & $-$             & 0.11\,$\pm$\,0.01 \\
9900 & $-$             & $-$             & 0.08\,$\pm$\,0.01 \\
\enddata
\end{deluxetable}

We also include previously reported measurements from the MOST 843\,MHz~\citep{1996A&AS..118..329W} and PMN 4850\,MHz~\citep{2007Ap&SS.307..423S} SNR catalogues. These surveys have intrinsic beam sizes of 45\arcsec\ and 4.3\arcmin, respectively, and do not resolve the PWN. 
The reported flux densities at these frequencies correspond to the integrated emission of the entire SNR.

\subsection{Radio Spectra of \snrname\ and \pwnname}
\label{sec:res-spectralIndex}

\begin{figure}[ht]
\includegraphics[trim=0 0 0 0,width=\columnwidth]{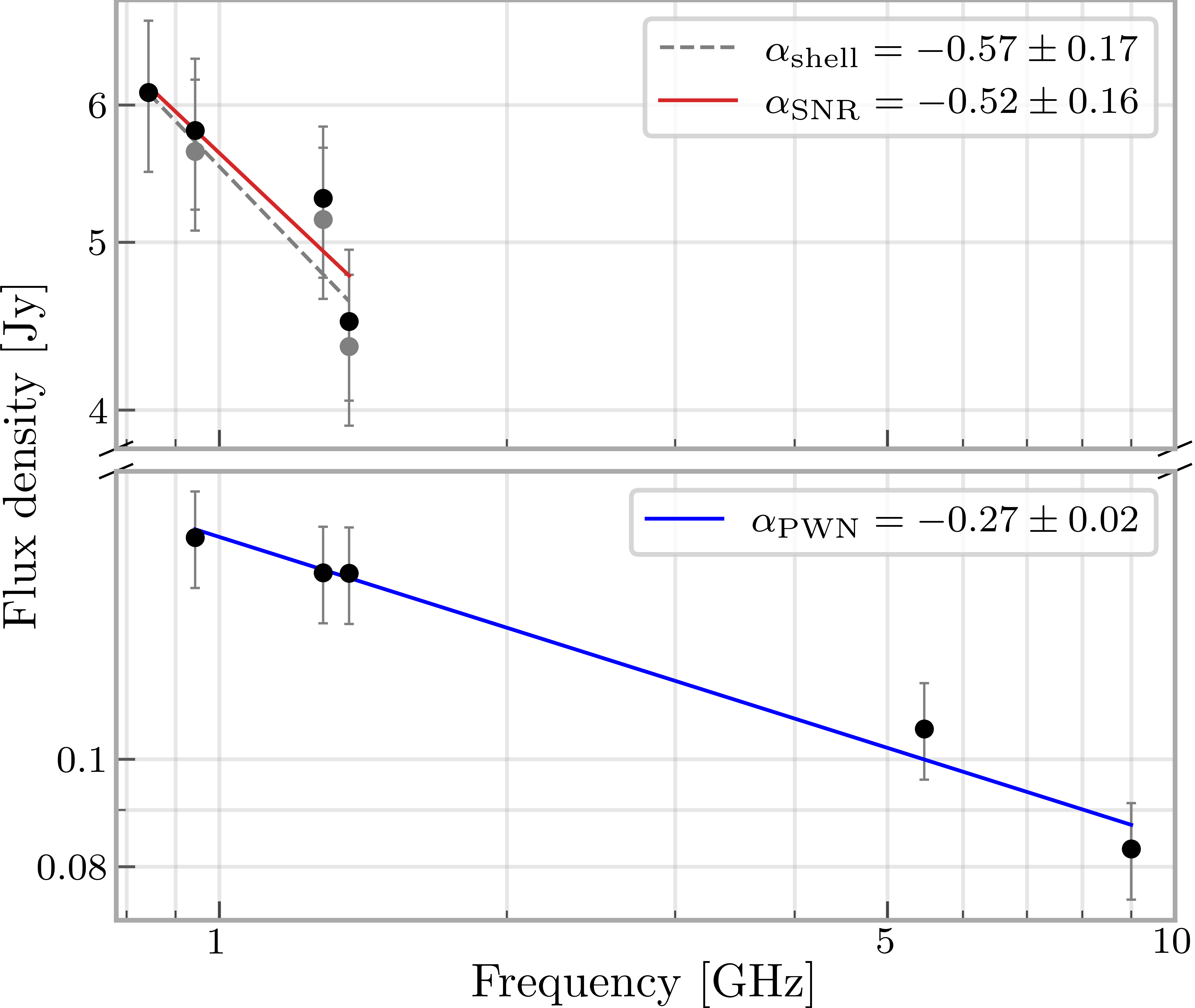}
\caption{Spectral index fits for \snrname$-$\pwnname\ system derived from the integrated flux densities listed in Table~\ref{tab:tab3-flux}, excluding the PMN point. A uniform 10\% uncertainty was assumed for all flux density measurements. Top: Spectral index fits for the entire SNR~(solid red line) and its shell~(dashed grey line). Bottom: Spectral index fit for the PWN~(solid blue line). Quoted uncertainties include only the statistical errors from the fit.}
\label{fig:fig4-spectralIndexGraph}
\end{figure}

\begin{figure*}
\includegraphics[trim=0 0 4 0, width=0.49\textwidth]{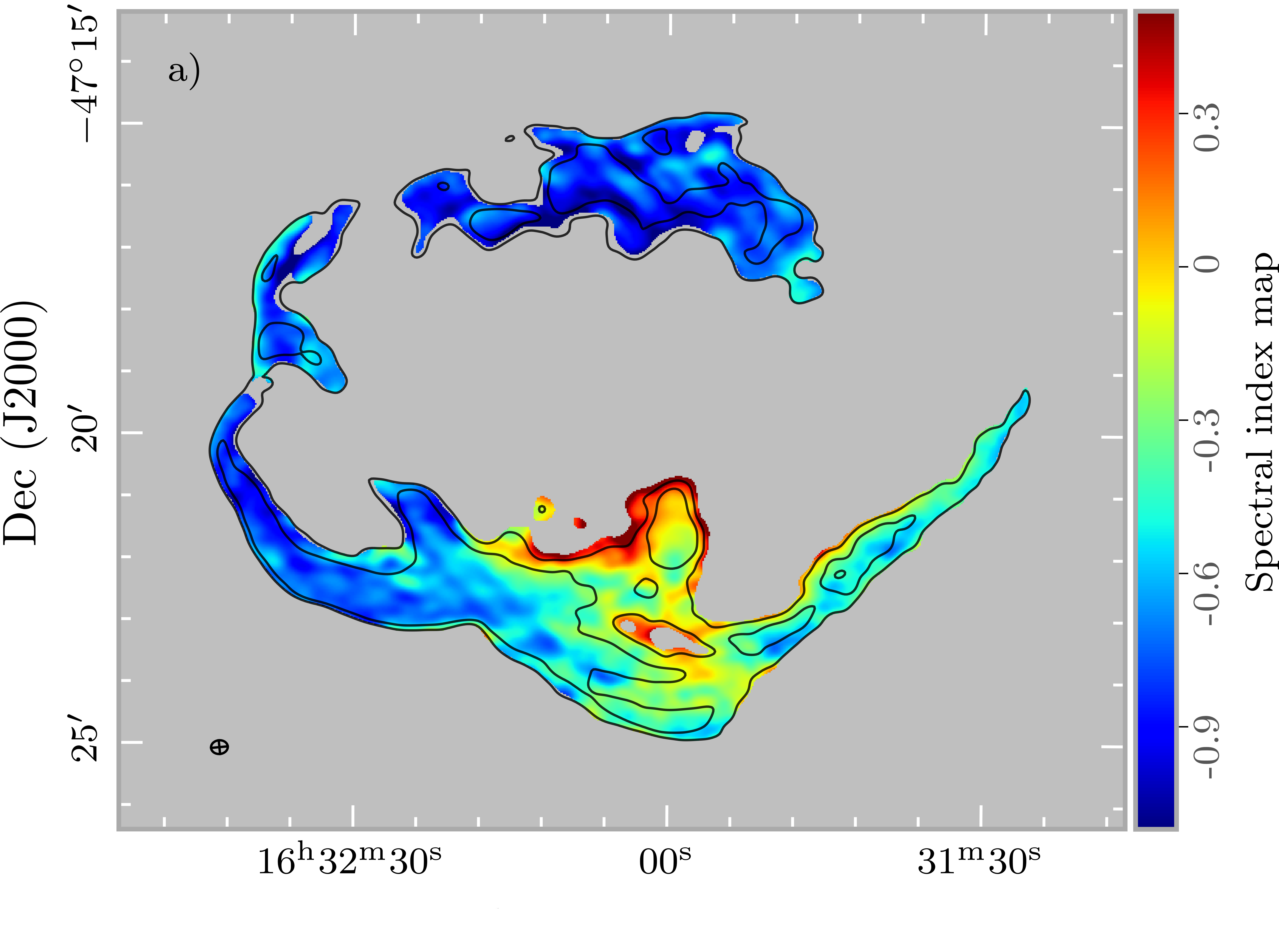}
\includegraphics[trim=4 0 0 0, width=0.49\textwidth]{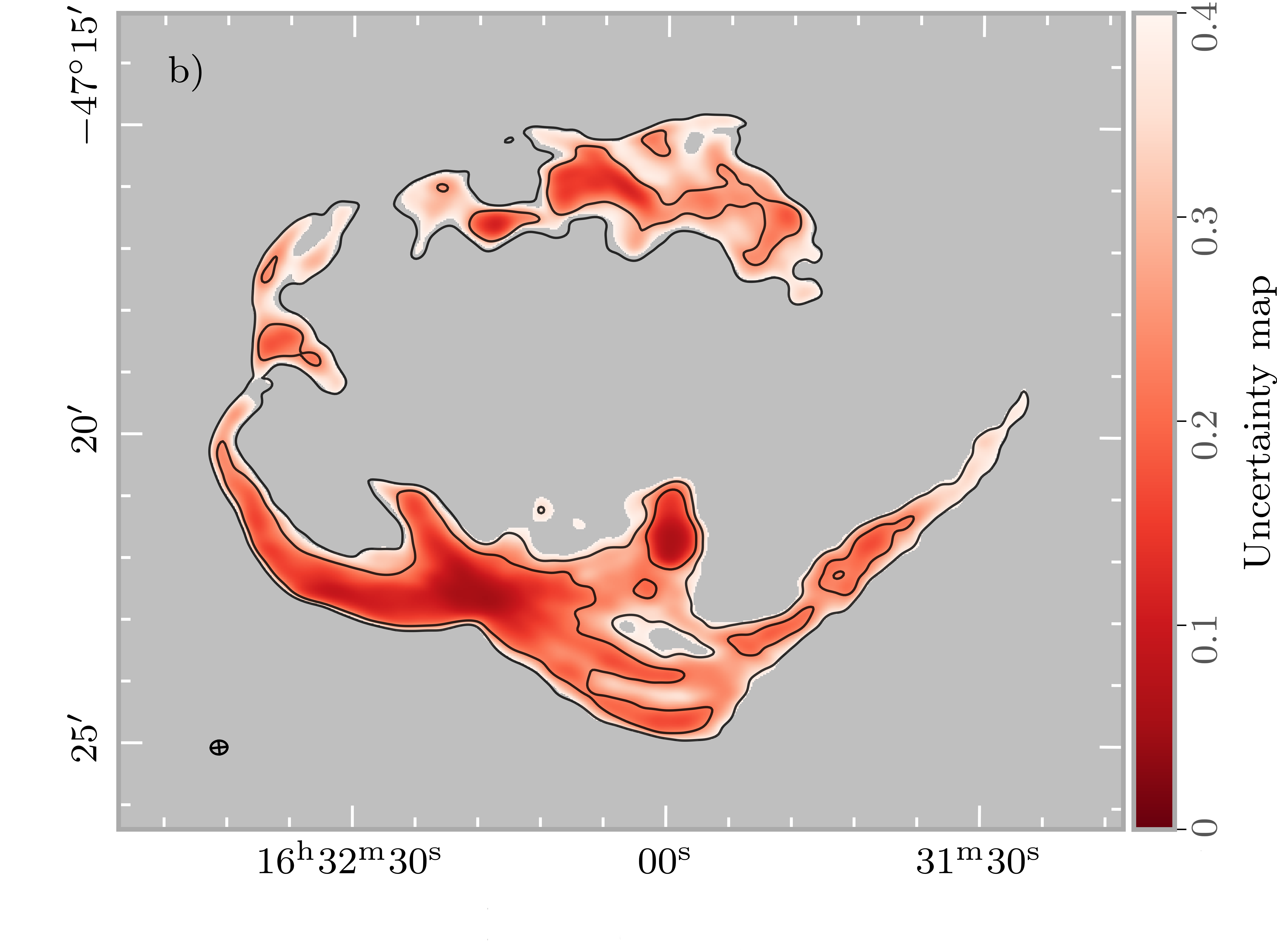}
\includegraphics[trim=0 0 3 0, width=0.33\textwidth]{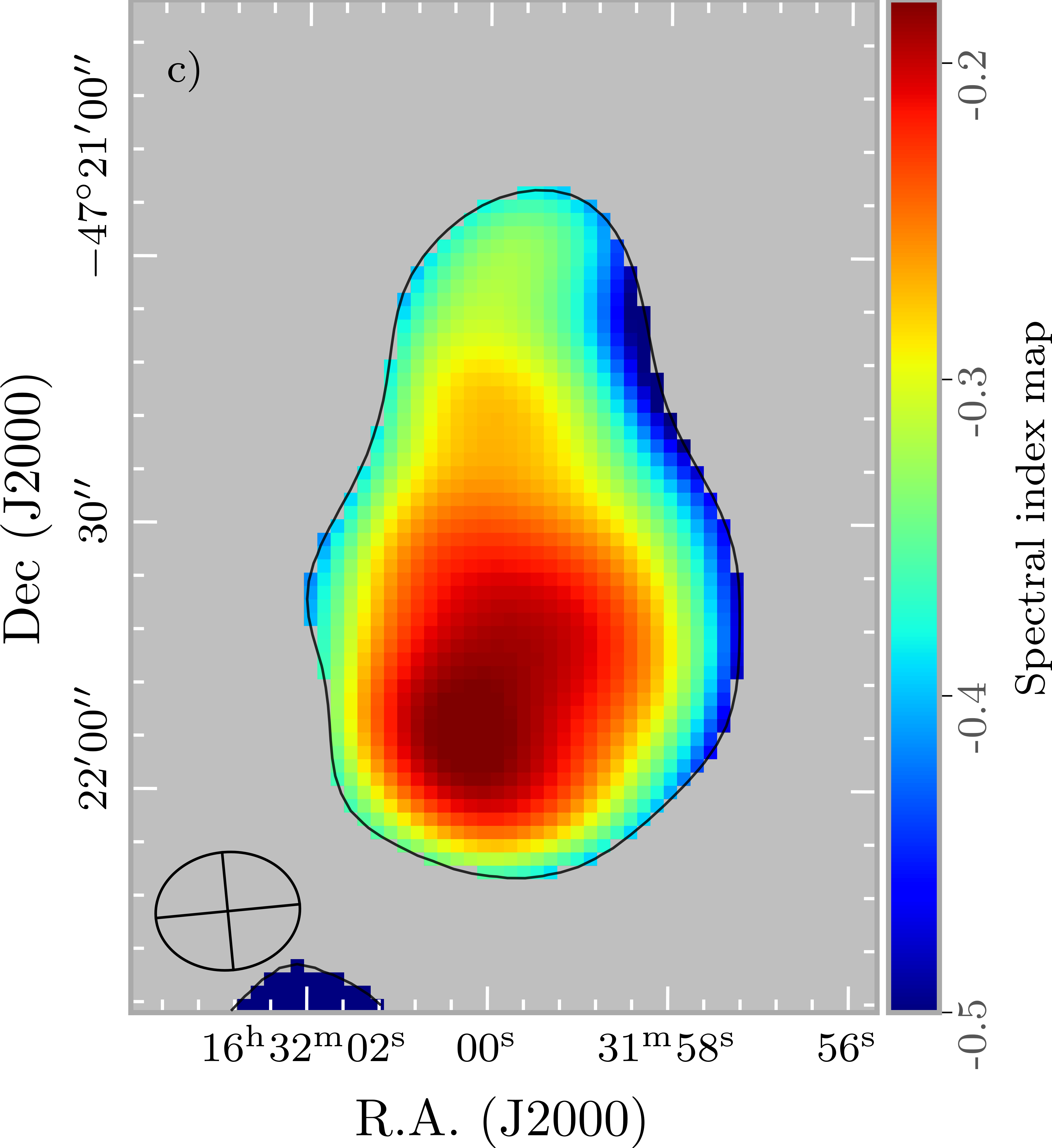}
\includegraphics[trim=1.5 0 1.5 0, width=0.33\textwidth]{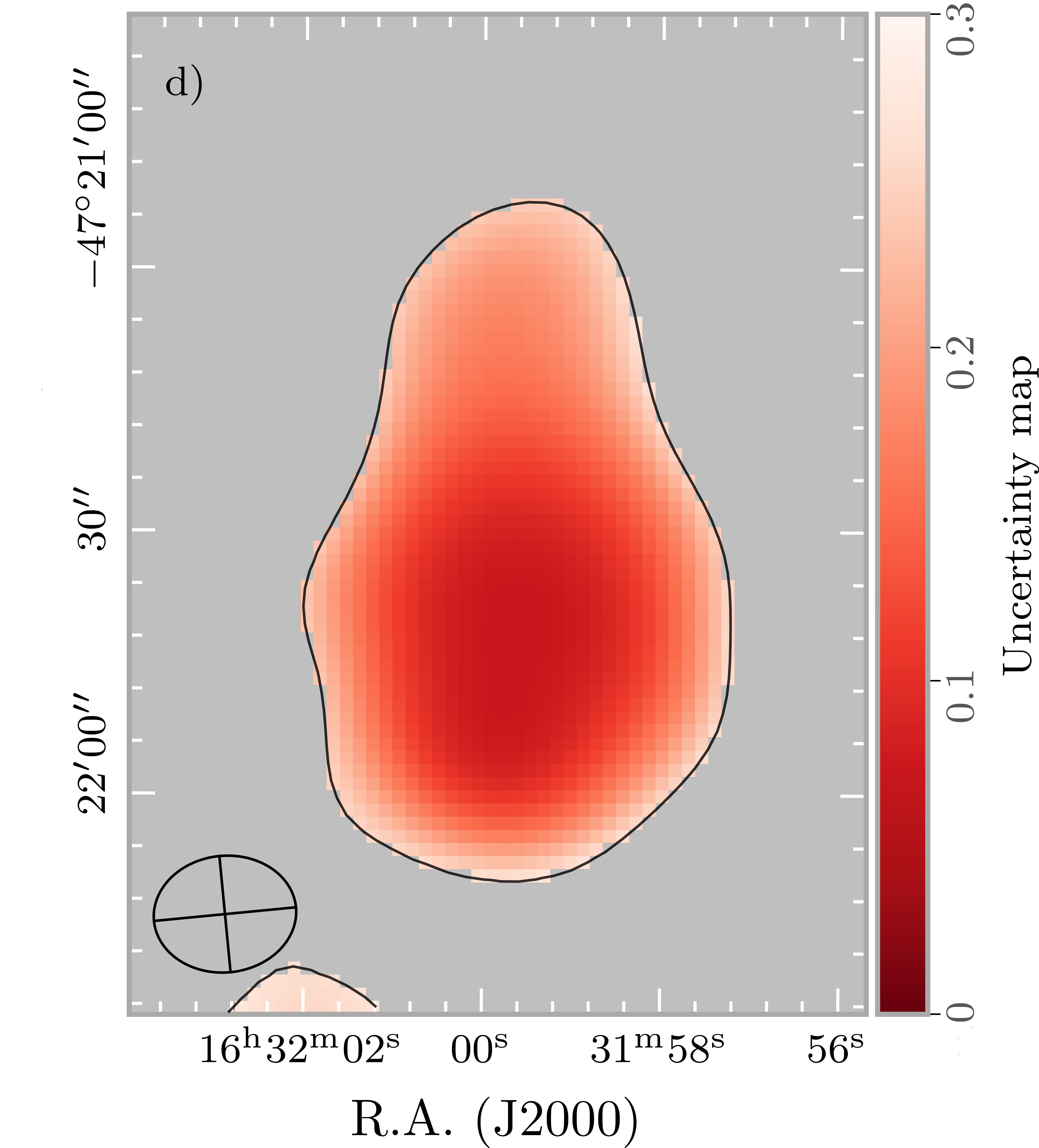}
\includegraphics[trim=3 0 0 0, width=0.33\textwidth]{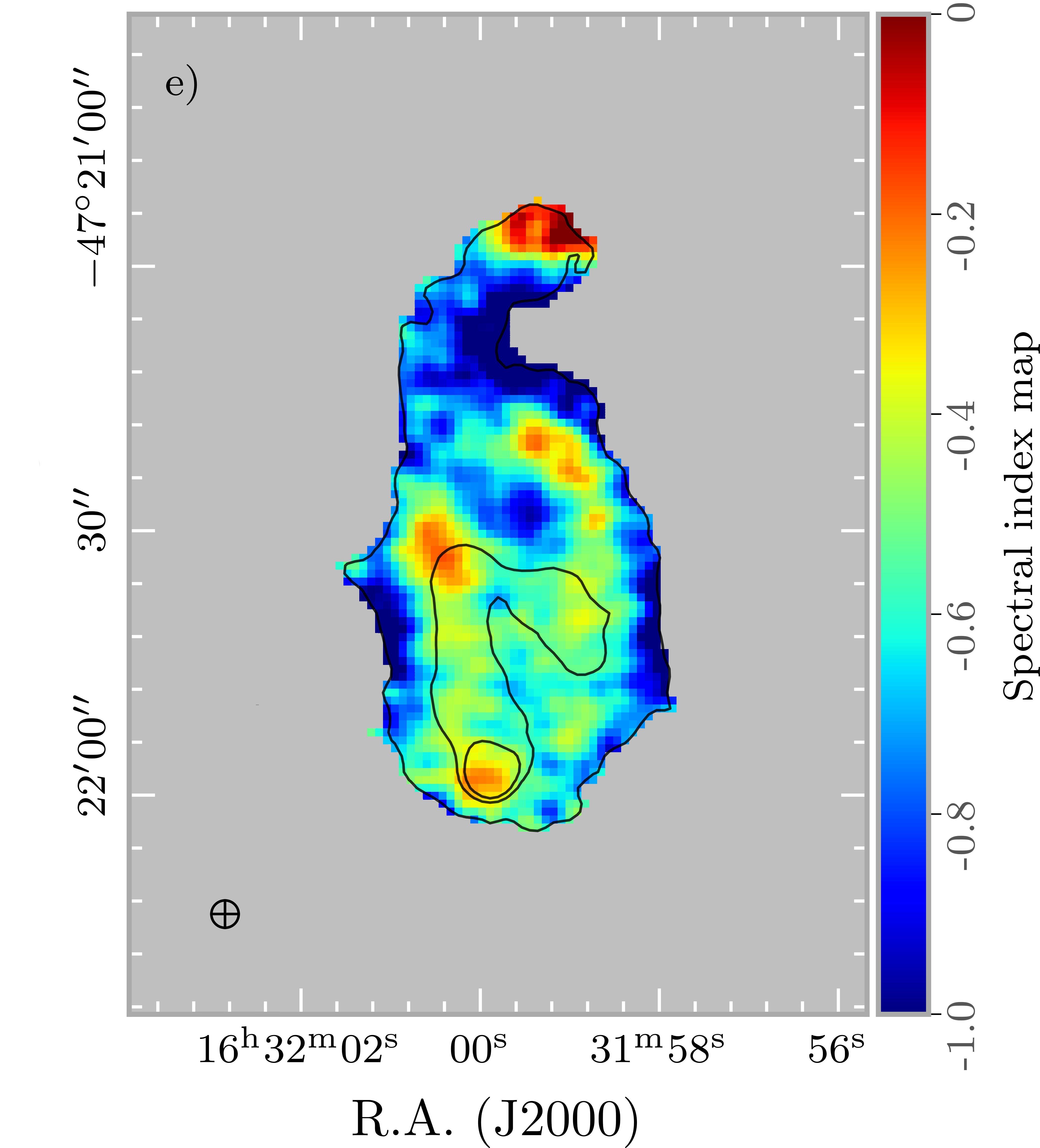}
\caption{Spectral index and corresponding uncertainty maps of \snrname\ SNR~(panels a and b) and \pwnname\ PWN~(panels c\,$-$\,e). The synthesised beam is shown as the black ellipse in the lower-left corner and is \ebs\ for all panels except~(e), where a 3\arcsec$\times$\,3\arcsec\ beam is used. Contour levels are identical to those shown in Figures~\ref{fig:fig2-snrAllFreq}~and~\ref{fig:fig3-pwnAllFreq}.}
\label{fig:fig5-spectralIndexMap}
\end{figure*}

SNRs and PWNe are predominantly non-thermal emitters, producing synchrotron radiation from radio to X-ray frequencies and beyond~\citep{book2}. In the radio band, this emission follows a power-law dependence of flux density on observing frequency, S$_{\nu}\propto\nu^{\alpha}$, where $\alpha$ is the spectral index. We use two approaches to examine the energy distribution of \snrname$-$\pwnname\ system. We calculate the spectral indices from integrated flux density measurements~(Table~\ref{tab:tab3-flux}; Figure~\ref{fig:fig4-spectralIndexGraph}), and generate spectral index maps 
(Figure~\ref{fig:fig5-spectralIndexMap}). The resulting spectral indices are summarised in Table~\ref{tab:tab4-spectralIndex}, including the value reported in previous studies for \snrname\ SNR of $\alpha=-0.50$~\citep[e.g.,][]{1970AuJPA..14..133S,2025JApA...46...14G}. 


In Figure~\ref{fig:fig4-spectralIndexGraph}, we present the spectral index fits derived separately for the entire SNR and its shell~(top panel), and for the PWN~(bottom panel). The fits were performed in log–log space using a simple linear least-squares regression\footnote{\href{https://docs.scipy.org/doc/scipy/reference/generated/scipy.stats.linregress.html}{SciPy \textsc{linregress}~\citep{2020NatMe..17..261V}}} applied to the measured flux densities. 
A uniform 10\% uncertainty was assumed for all measurements, including the MOST 843\,MHz point. The PMN measurement was excluded from the fitting as it represents a significant outlier. 
This discrepancy is most likely related to the large beam size of the PMN survey, which may complicate the separation of the relatively faint SNR emission from the surrounding Galactic background. 
After excluding PMN, we calculated spectral indices of $\alpha_{\rm SNR}$\,=\,$-0.52\pm0.16$ for the entire SNR, $\alpha_{\rm shell}$\,=\,$-0.57\pm0.17$ for the shell, and $\alpha_{\rm PWN}$\,=\,$-0.27\pm0.02$ for the PWN. The given uncertainties represent the statistical errors of the linear regression. 

Since the formal statistical errors likely underestimate the true uncertainty, we further assessed the robustness of the spectral indices using a bootstrap approach, following the methodology of \citet{2025ApJ...980..162J}. We generated 1000 realisations of the spectrum by perturbing each flux density according to its uncertainty and refitting a power law. The resulting distributions of spectral indices are well described by Gaussians, and their standard deviations are adopted as the final uncertainties given in Table~\ref{tab:tab4-spectralIndex}. 

Spatially resolved spectral index maps were produced for \pwnname\ using  ASKAP, MeerKAT and ATCA data, while for \snrname\ we exclude the ATCA observations because of the insufficient FoV of a single pointing. As described in Section~\ref{sec:res-flux}, all images were regridded to a common pixel scale of 1.5\arcsec$\times$\,1.5\arcsec\ and then convolved to a common angular resolution of \ebs. 
Spectral index maps were generated using a weighted linear regression of the logarithmic flux density as a function of logarithmic frequency. The weights were determined from the rms noise in each image. The slope of the best fit through the data at each pixel was used to construct the spectral index map(Figures~\ref{fig:fig5-spectralIndexMap}a,~\ref{fig:fig5-spectralIndexMap}c~and~\ref{fig:fig5-spectralIndexMap}e). Pixels with flux densities below the 5$\sigma$ threshold in at least one frequency band were excluded from the fit and set to NaN values, appearing as grey 
in the final maps. 

We also generated a spectral index uncertainty map using standard error propagation, following a procedure similar to that described by \citet{2023A&A...672A..28L,2025SerAJ.210...13S}. For each frequency pair~($\nu_1,\nu_2$), the uncertainty was calculated as $\sigma_{\alpha} = [\ln(\nu_1 / \nu_2)]^{-1}\sqrt{(\sigma_1/S_1)^2 + (\sigma_2/S_2)^2}$, where $S_i$ and $\sigma_i$ are the flux density and rms noise of each image. This calculation was repeated for all frequency pairs, and the final uncertainty map was obtained by averaging the individual pairwise uncertainty maps~(Figures~\ref{fig:fig5-spectralIndexMap}b~and~\ref{fig:fig5-spectralIndexMap}d).

The spectral index map of \snrname\ reveals significant spatial variability, with the northern diffuse region exhibiting systematically steeper spectra than the southern side of the remnant. Typical uncertainties range from $\sim$\,0.10$-$0.25 across the bright shell and increase towards the fainter outer regions. Pixels with uncertainties exceeding $\sigma$\,$>$\,0.4 were masked and excluded from further analysis. 
From the remaining pixels we measure average spectral indices of $\alpha_{\rm SNR}$\,=\,$-$0.53\,$\pm$\,0.25 for the entire SNR and $\alpha_{\rm shell}$\,=\,$-$0.56\,$\pm$\,0.26 for the shell. 
%
The spectral estimates for \pwnname\ were restricted to the same region used for the integrated flux density measurements~(Figures~\ref{fig:fig2-snrAllFreq}a~and~\ref{fig:fig3-pwnAllFreq}a). We derive an average spectral index of $\alpha_{\rm PWN}=-0.28\pm0.15$. 

The spectrum of the whole SNR is flatter than that of the shell alone, as expected from the contribution of the flat-spectrum PWN to the total emission~(Table~\ref{tab:tab4-spectralIndex}). The close agreement between the results obtained using independent approaches, integrated spectral fitting and spectral index mapping, indicates that the observed trends reflect intrinsic source properties rather than methodological or imaging systematics. Our results for \snrname\ are consistent with the previously reported spectral index and the canonical value of $-$0.5~\citep{Bell1978}. 

\begin{deluxetable}{l c c c}
\tablecaption{Radio spectral indices and X-ray photon index of \snrname$-$\pwnname\ system. The literature spectral index is taken from \citet{1970AuJPA..14..133S}. Spectral indices derived in this work are obtained from integrated spectral fitting and spectral index mapping. The photon index is obtained from \xmm\ spectroscopy~(Section~\ref{sec:res-x-ray}).}
\label{tab:tab4-spectralIndex}
\tablehead{
\colhead{} & \colhead{\snrname} & \colhead{\snrname\ shell} & \colhead{\pwnname}
}
\startdata
$\alpha_{\rm literature}$ & $-$0.50              & $-$                  & $-$\\
$\alpha_{\rm fit}$      & $-$0.52\,$\pm$\,0.25 & $-$0.57\,$\pm$\,0.25 & $-$0.27\,$\pm$\,0.05\\
$\alpha_{\rm map}$        & $-$0.53\,$\pm$\,0.25 & $-$0.56\,$\pm$\,0.26 & $-$0.28\,$\pm$\,0.15\\
\hline
$\Gamma_{\rm \xmm}$       & $-$                  & $-$                  & 1.6\,$\pm$\,0.4 \\
\enddata
\end{deluxetable}

The ATCA 3\arcsec$\times$\,3\arcsec\ total power images of \pwnname~(Figures~\ref{fig:fig3-pwnAllFreq}d~and~\ref{fig:fig3-pwnAllFreq}e) reveal compact brightness enhancements~(knots) that are not visible in the lower resolution maps. To investigate whether these features are associated with spectral structure, we produced a matched-resolution two-point spectral index map~(Figure~\ref{fig:fig5-spectralIndexMap}e). The resulting map shows a flatter spectrum coincident with the radio apex peak, whereas only one of the bright knots has a possibly corresponding flatter spectral patch. The spectral flattening toward the end of the tail is likely dominated by increased noise at low surface brightness. Additional small-scale flattening on the western side may similarly reflect imaging or sensitivity effects rather than intrinsic spectral structure. Overall, the spectral map suggests a locally flatter spectrum at the radio peak, while the compact knots do not exhibit clearly corresponding spectral features.

\subsection{Polarisation Properties of \pwnname}
\label{sec:res-polarisation}

The pulsar wind, composed of relativistic particles propagating through the ambient magnetic field, produces intrinsically linearly polarised synchrotron emission~\citep{1956BAN....12..285O,2006ARA&A..44...17G}. Because the emitting plasma is magnetised and composed of relativistic charged particles, particle motion is strongly coupled to the 
nebular magnetic field, making radio polarimetry a powerful probe of 
its geometry. The distribution of polarised intensity and fractional polarisation provides direct information on the level of magnetic field ordering, spatial variations can trace turbulence, compression, or depolarisation 
along the line of sight~\citep{2006ARA&A..44...17G,2017ASSL..446....1K}. 

\begin{figure*}[ht]
\includegraphics[trim=0 0 3 0, width=0.33\textwidth]{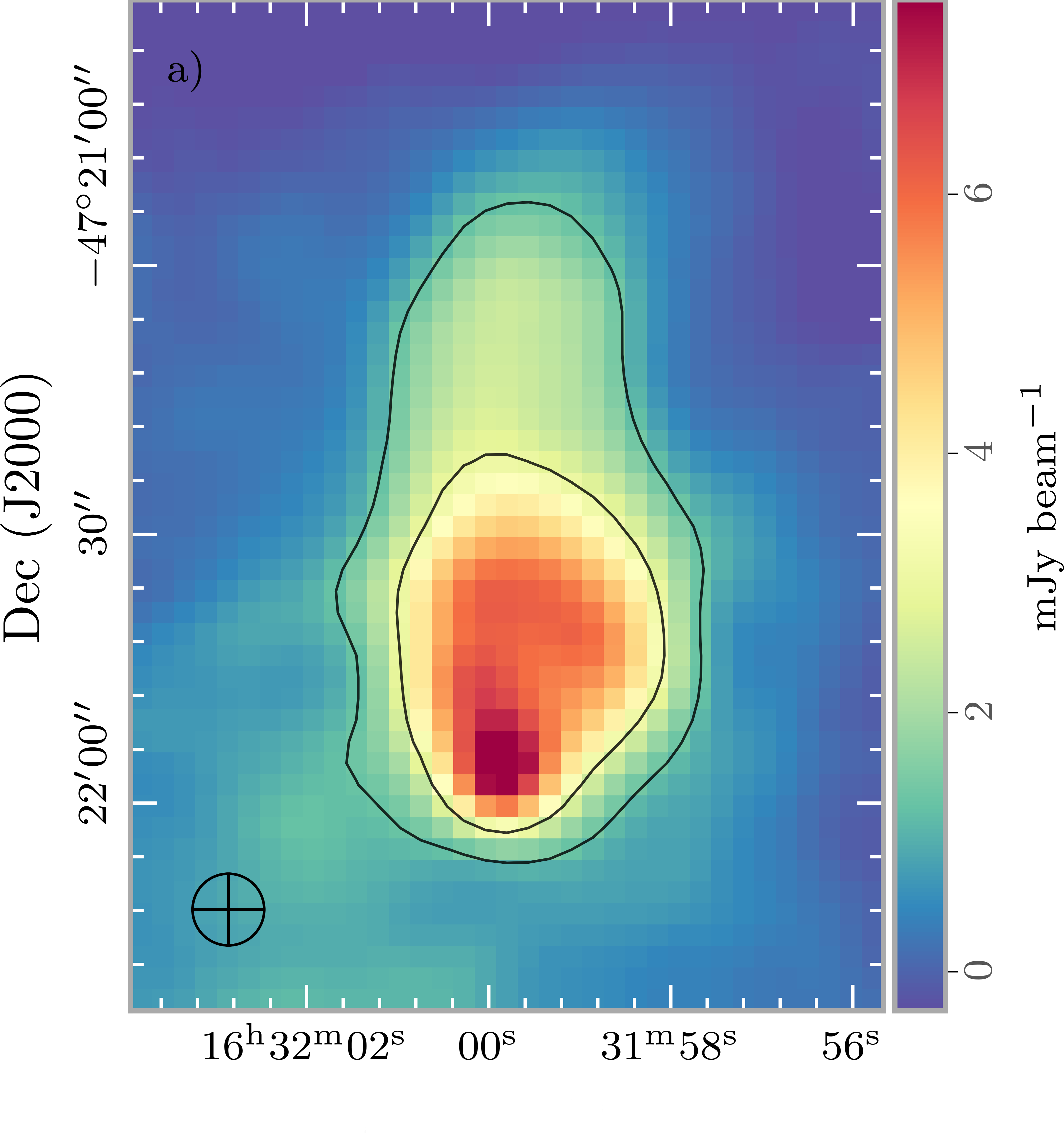}
\includegraphics[trim=1.5 0 1.5 0, width=0.33\textwidth]{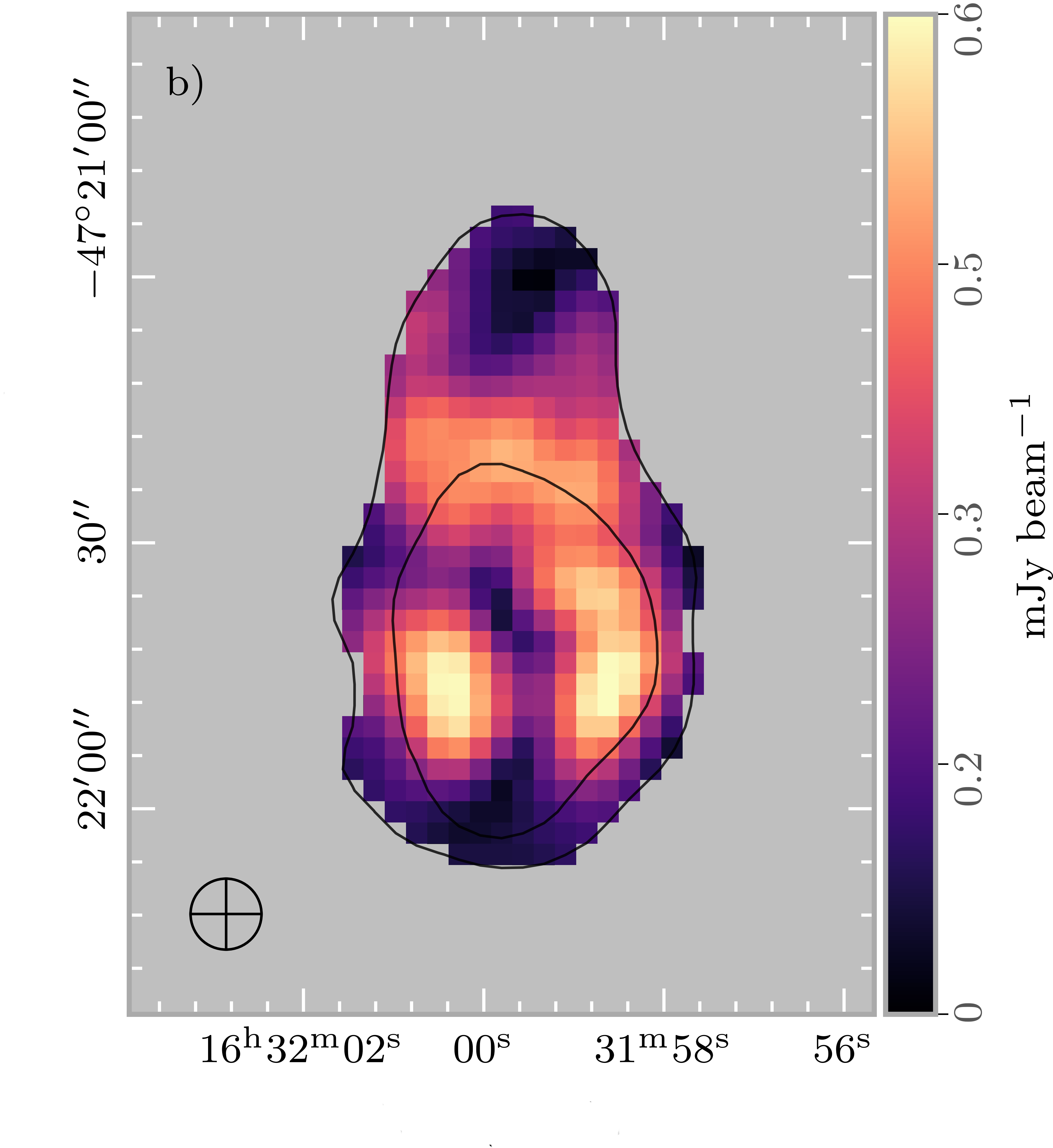}
\includegraphics[trim=3 0 0 0, width=0.33\textwidth]{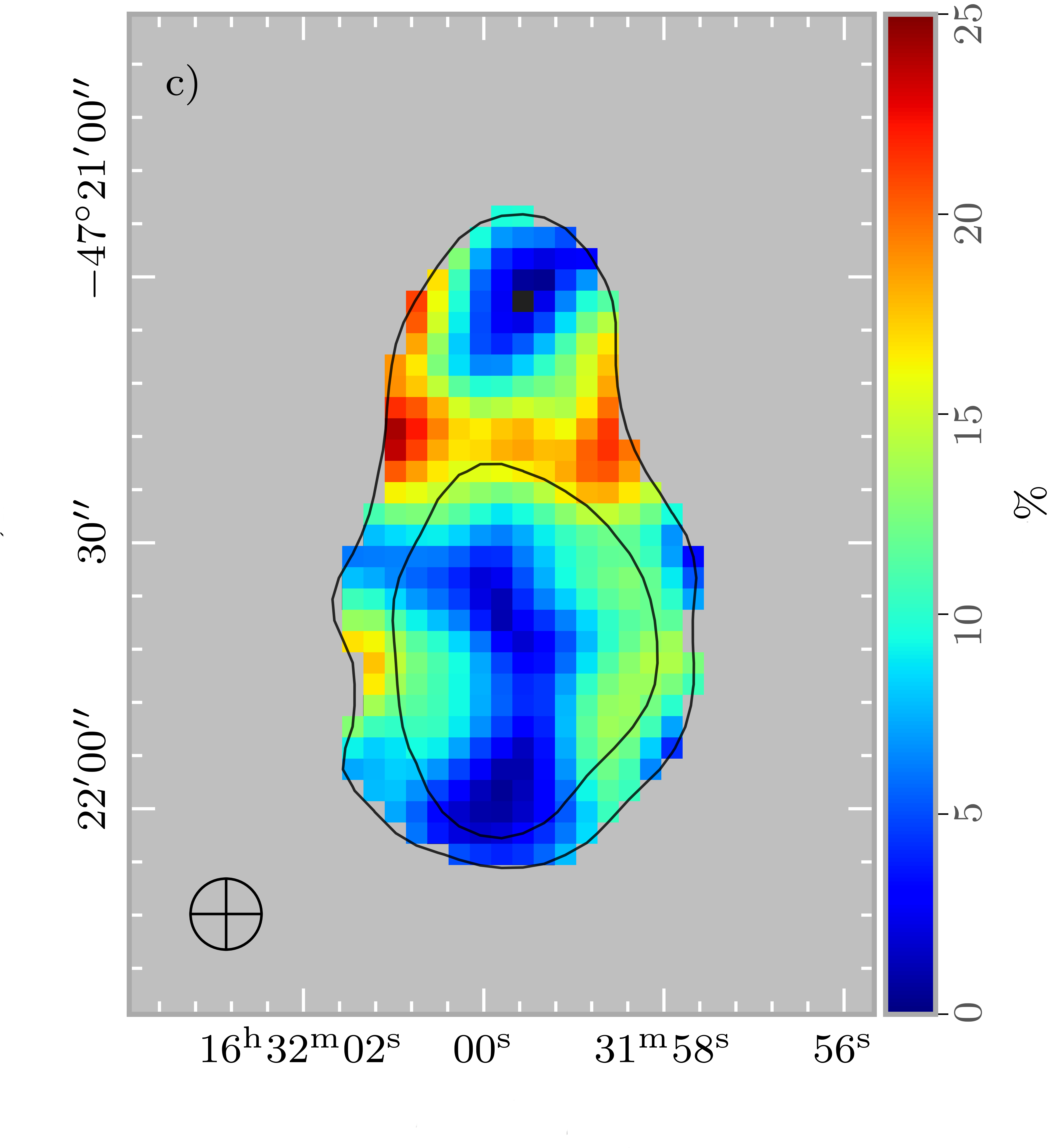}
\includegraphics[trim=0 0 3 0, width=0.33\textwidth]{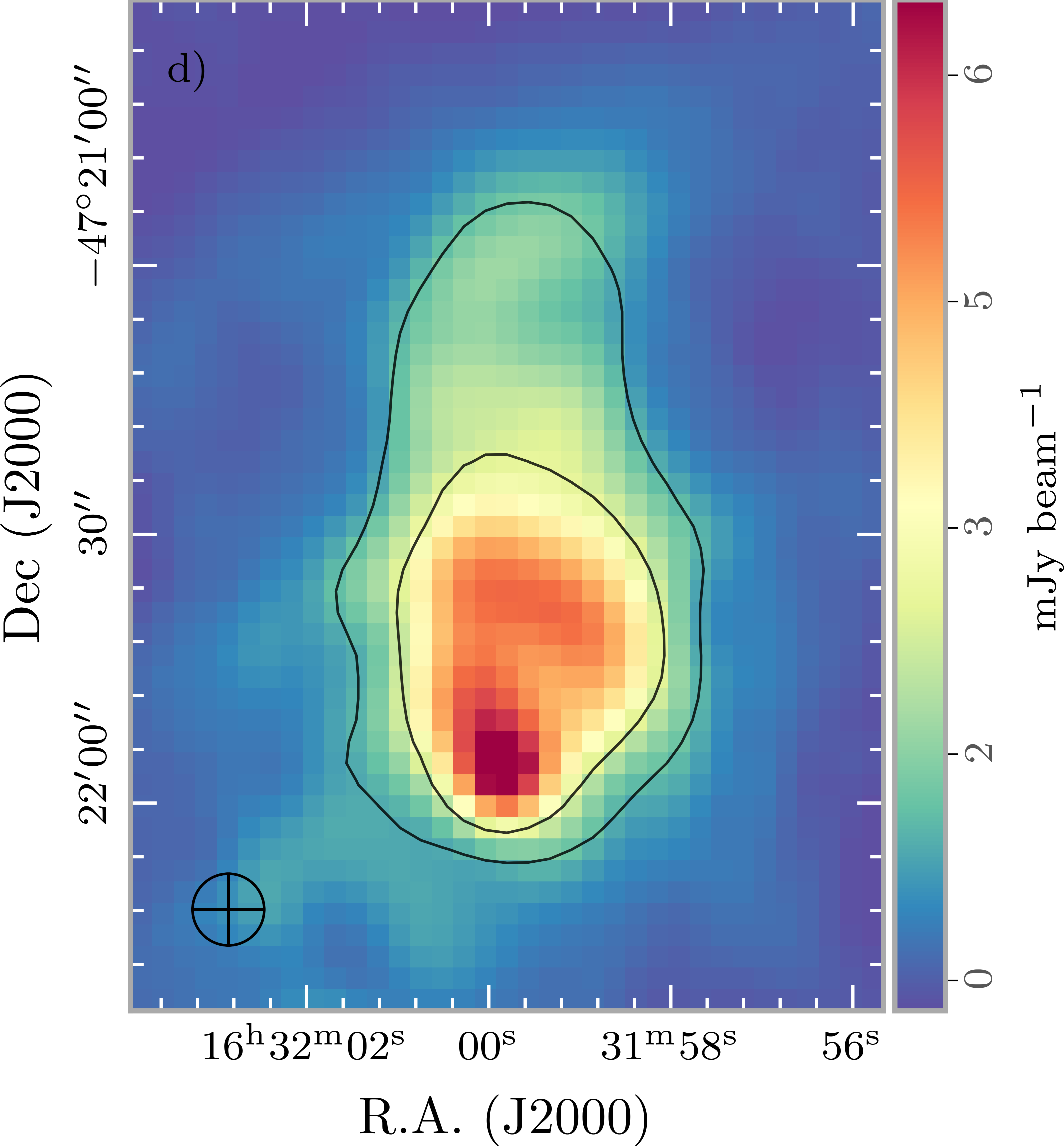}
\includegraphics[trim=1.5 0 1.5 0, width=0.33\textwidth]{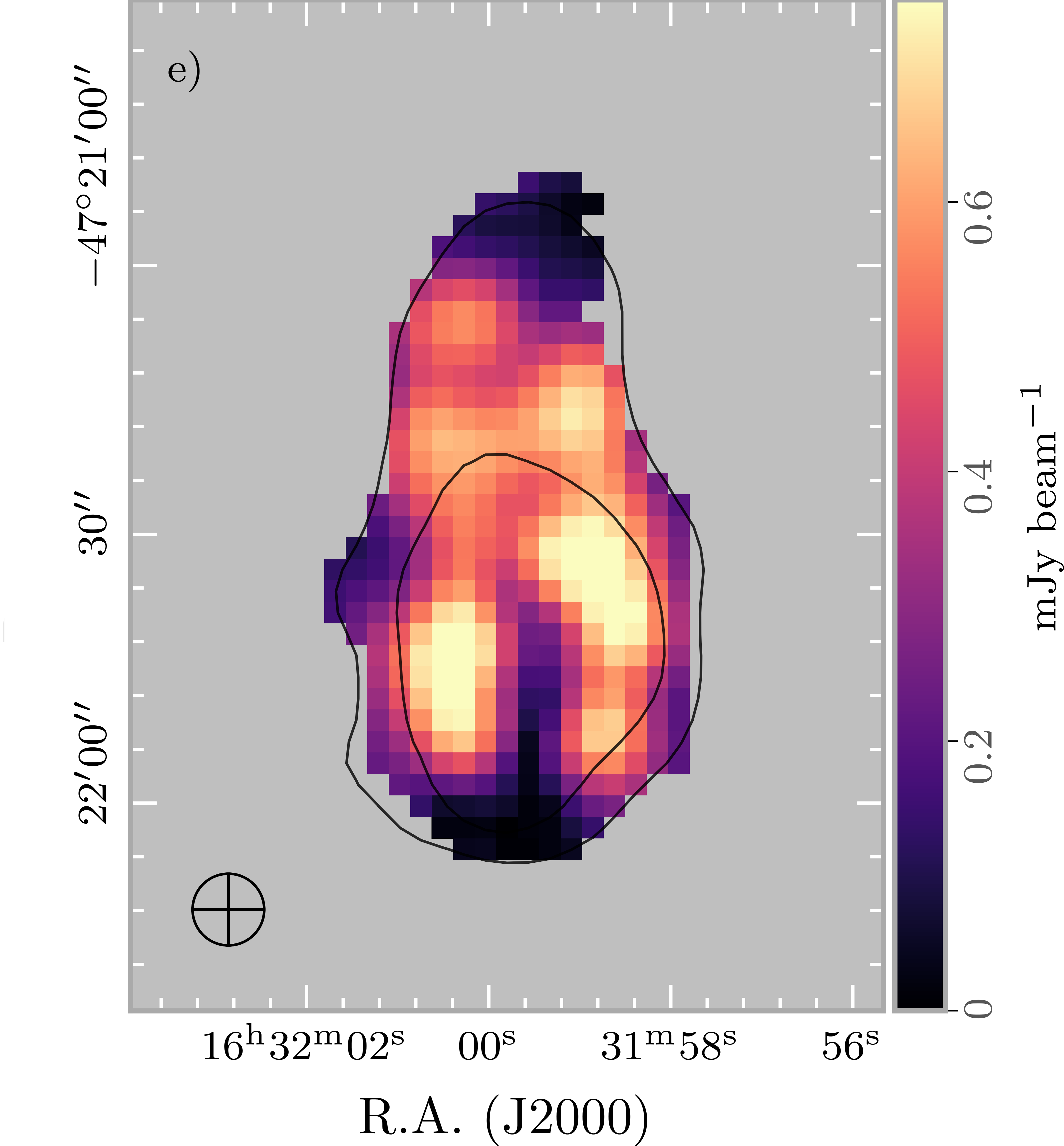}
\includegraphics[trim=3 0 0 0, width=0.33\textwidth]{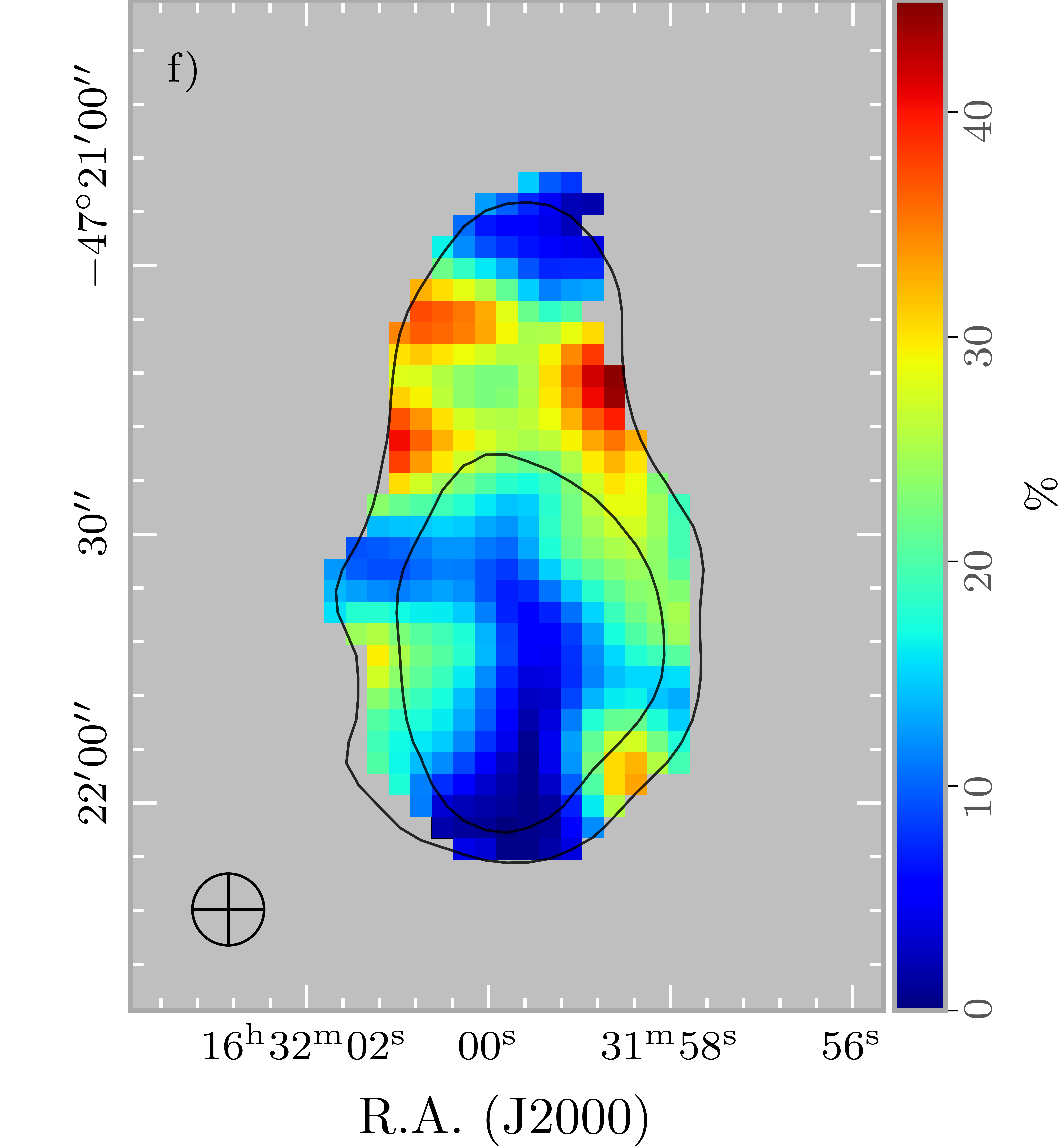}
\caption{ATCA total intensity~(left), polarised intensity~(middle), and fractional polarisation~(right) maps of \pwnname\ at \al~(panels a\,$-$\,c) and \ah~(panels d\,$-$\,f). The black circle in the lower-left corner indicates the synthesised beam of 6\arcsec$\times$\,6\arcsec. Black contours correspond to the \al\ total intensity image at levels of 1.3 and 3\,\mjb.}
\label{fig:fig6-polarisation}
\end{figure*}
\begin{figure*}
\includegraphics[trim=0 0 4 0, width=0.33\textwidth]{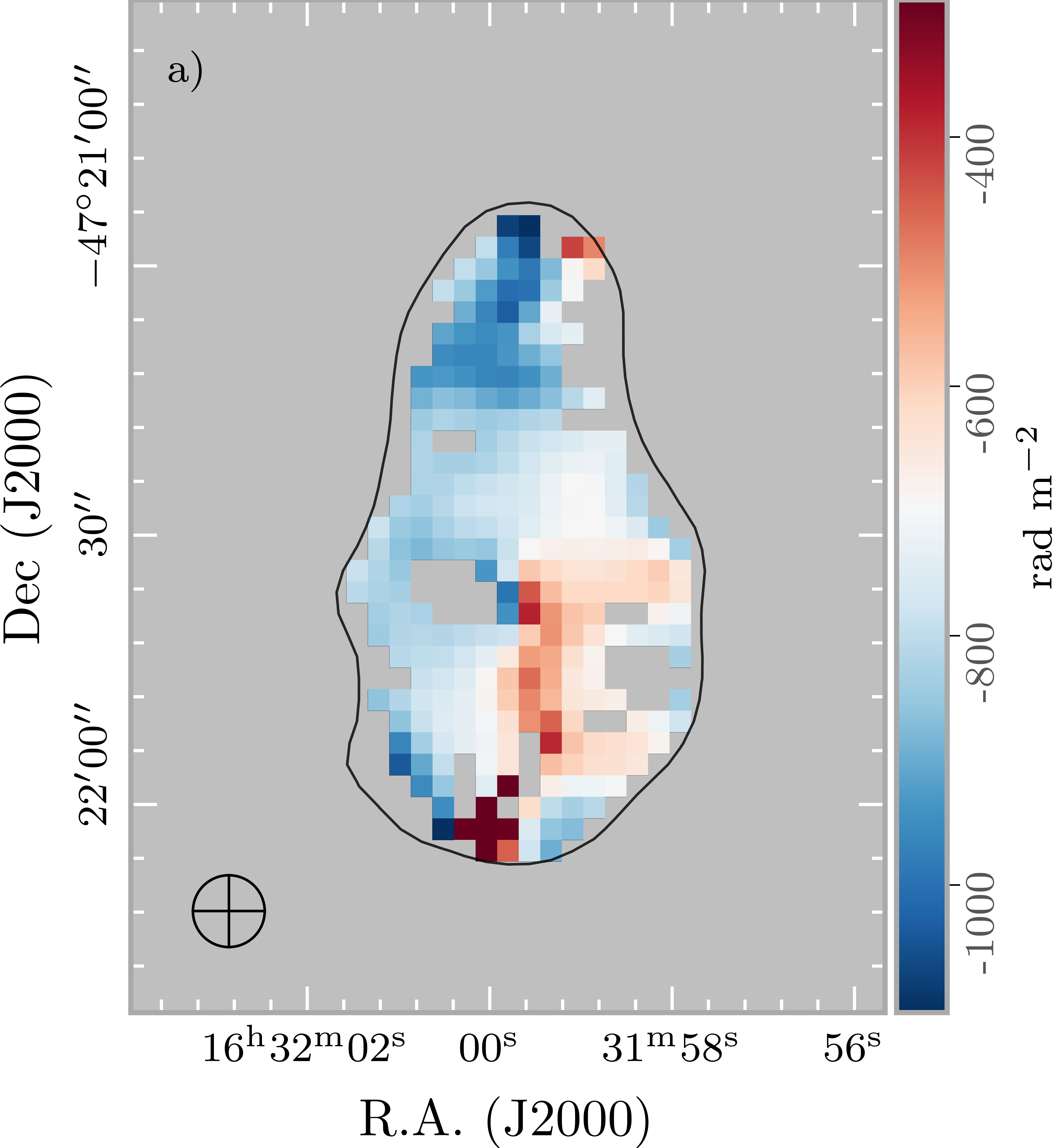}
\includegraphics[trim=2 0 2 0, width=0.33\textwidth]{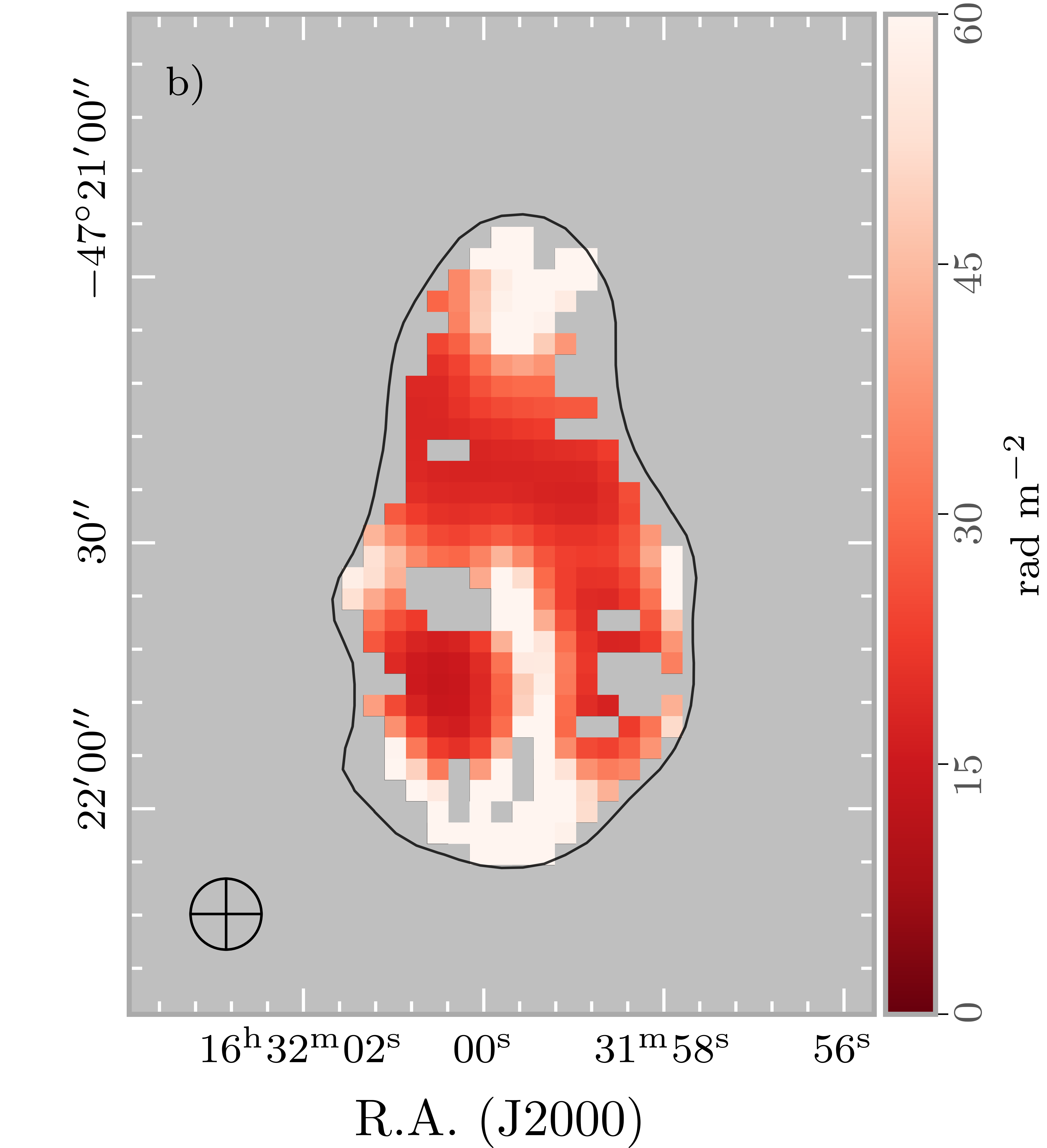}
\includegraphics[trim=4 0 4 0, width=0.33\textwidth]{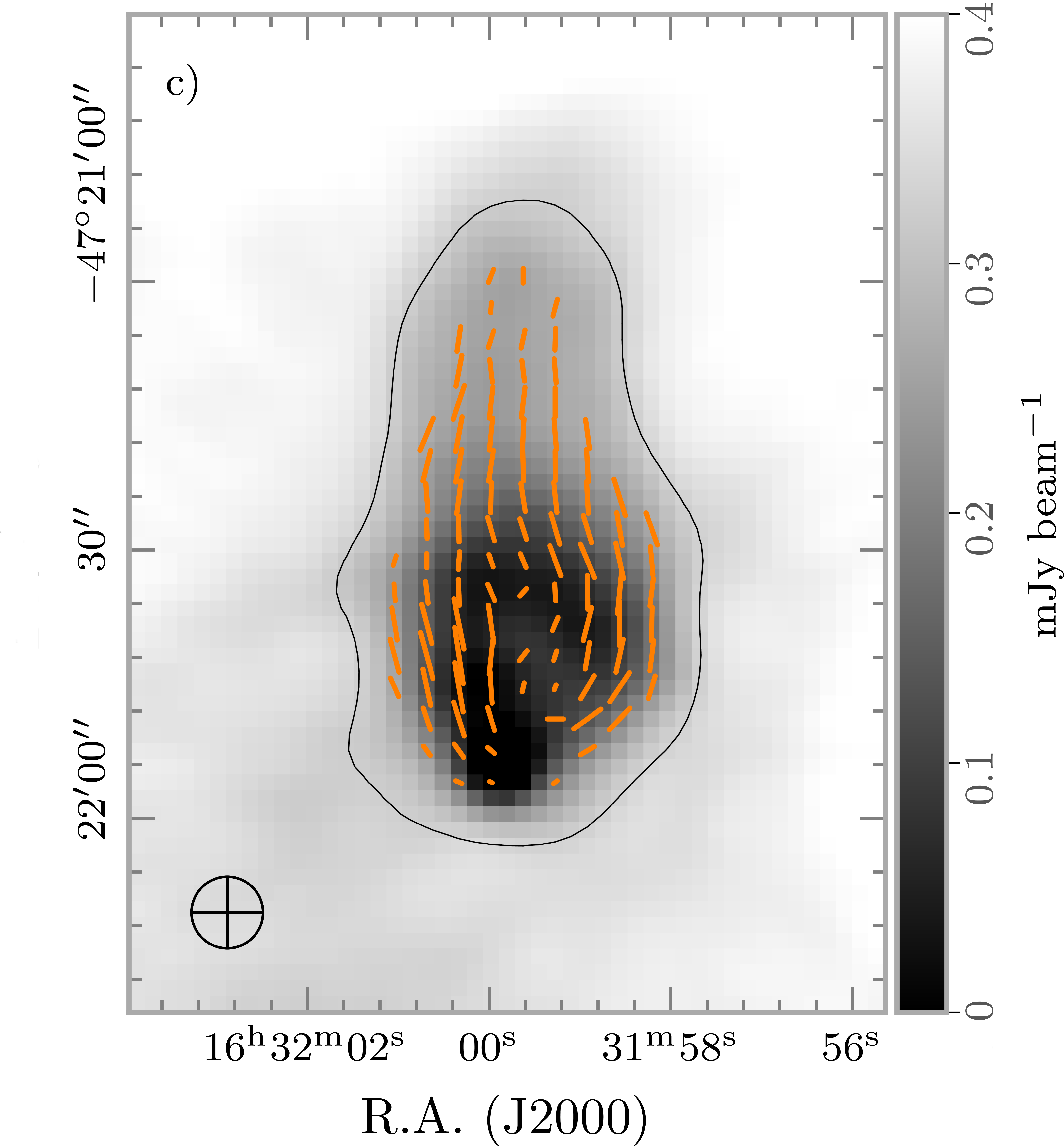}
\caption{RM map of \pwnname\ (a), corresponding RM uncertainty map (b), and projected magnetic field vectors overlaid on the ATCA \al\ total intensity image (c). Vector lengths are proportional to the fractional polarisation shown in Figure~\ref{fig:fig6-polarisation}c. The circle in the lower-left corner indicates the synthesised beam of 6\arcsec$\times$\,6\arcsec. The contour corresponds to the 
\al\ total intensity 
level of 1.3\,\mjb.}
\label{fig:fig7-rm}
\end{figure*}

For the polarisation analysis of \pwnname, we used ATCA observations at 5500 and \ah\ obtained in full Stokes parameters. All Stokes images were restored with a common beam of 6\arcsec$\times$\,6\arcsec. The linearly polarised intensity was calculated as $P=\sqrt{Q^{2}+U^{2}}$, where $Q$ and $U$ are the flux densities of the corresponding Stokes parameters. The fractional polarisation was calculated as $p=P/I$, where $I$ is the Stokes\,$I$ flux density. 
The resulting total intensity, polarised intensity, and fractional polarisation maps at both frequencies are shown in Figure~\ref{fig:fig6-polarisation}. 

Significant linearly polarised emission is detected across most of the nebula 
(Figures~\ref{fig:fig6-polarisation}b~and~\ref{fig:fig6-polarisation}e). The polarised intensity distribution broadly follows the overall PWN morphology but shows clear spatial variations that do not directly trace the total intensity structure. The peak of the Stokes\,$I$ emission does not coincide with a maximum in polarised intensity. Instead, a low polarised intensity is observed near the intensity peak, which extends through the middle towards the northern part of the nebula. Regions of enhanced polarised intensity do not coincide with local maxima of the total intensity. In the tail beyond the bright part, the polarised intensity remains relatively high despite the significant decline in total intensity.

The fractional polarisation maps are shown in Figures~\ref{fig:fig6-polarisation}c~and~\ref{fig:fig6-polarisation}f. We measure an average fractional polarisation of 10$\pm$3\% at \al\ and 18$\pm$4\% at \ah. The highest fractional polarisation values are observed in the fainter outer regions of the tail. Although this enhancement is partly associated with a decline in Stokes\,$I$, the polarised intensity itself also increases in this region, reaching up to 18$\pm$1\% at \al\ and 29$\pm$3\% at \ah. 
Pixels with extreme fractional polarisation values, typically occurring where the total emission is faint and uncertainties are large, near the nebular edges, were excluded from the analysis. 
No significant emission was detected in Stokes\,$V$ at either 5500 or \ah\ above the 
5$\sigma$ level, and no evidence for circular polarisation associated with \psr\ was found.

\subsection{Rotation measure distribution across \pwnname}
\label{sec:res-rm}

As linearly polarised radiation propagates through a magneto-ionised ISM, its polarisation angle is rotated by free thermal electrons embedded in a magnetic field. 
This effect, known as Faraday rotation~\citep{1966MNRAS.133...67B}, is described by the rotation measure~(RM) relation $\phi_\lambda$\,=\,$\phi_0+$RM\,$\lambda^2$, where $\phi_\lambda$ is the observed polarisation angle at wavelength $\lambda$, and $\phi_0$ is the intrinsic polarisation angle. The RM can be quantified using multi-frequency observations~\citep[][for a review]{2021MNRAS.507.4968F} and provides a probe of magnetised plasma along the line of sight. 
After correcting the observed polarisation angles for Faraday rotation, the intrinsic orientation of the projected magnetic field in the source can be recovered.

To derive the RM map of \pwnname, we divided the 2048\,MHz bandwidth of the ATCA \al\ observation into four 512\,MHz sub-bands centred at 4732, 5244, 5756, and 6268\,MHz. For each sub-band, Stokes $Q$ and $U$ maps were produced following the same procedure used for the full-band images~(see Section~\ref{sec:atca}) and restored using a common synthesised beam of 6\arcsec$\times$\,6\arcsec. Polarisation angles were calculated from the Stokes~$Q$ and $U$ images at each frequency, and a linear least-squares fit of $\phi(\lambda^2)$ was performed on a pixel-by-pixel basis to generate the RM map~(Figure~\ref{fig:fig7-rm}a). The RM uncertainty map~(Figure~\ref{fig:fig7-rm}b) was derived from the 1$\sigma$ uncertainties of the fitted slopes. The n$\pi$ ambiguity was resolved by enforcing continuity of $\phi(\lambda^2)$ between adjacent frequency channels prior to fitting. Only pixels exceeding a 5$\sigma$ threshold in both Stokes~$Q$ and $U$ were included in the analysis to ensure reliable polarisation angle determination. 

The resulting RM map traces the spatial distribution of Faraday rotation across \pwnname. The RM values are predominantly negative over the entire nebula, ranging from approximately $-$300 to $-$1000\,rad\,m$^{-2}$, with a weighted mean of $-752$ $\pm$ 31\,rad\,m$^{-2}$. A clear east-west gradient is observed, with more negative values on the eastern side~(average $-817$\,$\pm$\,32\,rad\,m$^{-2}$) compared to the western side~($-685$\,$\pm$\,30\,rad\,m$^{-2}$). The middle region shows smaller RM magnitudes of about $-517$\,rad\,m$^{-2}$, although with increased uncertainty~(Figure~\ref{fig:fig7-rm}b). 
A small number of pixels near the tip of \pwnname\ have positive RM values, reaching up to +800\,rad\,m$^{-2}$, with an average of +467\,$\pm$\,210\,rad\,m$^{-2}$. However, these appear as isolated patches with large uncertainties and do not form coherent structures. Given the limited frequency sampling and low polarised signal-to-noise ratio in these pixels, this feature is most likely due to unstable or wrapped fits rather than true magnetic field reversals, and is therefore not considered physically significant.

The uncertainty map closely follows the distribution of polarised intensity~(Figure~\ref{fig:fig6-polarisation}b), with the lowest uncertainties found in regions of bright polarised emission and larger uncertainties in fainter parts of the nebula. This indicates that the RM measurements tracing the global east-west gradient are robust, while fine-scale features in low-signal regions should be interpreted with caution.

The observed polarisation angles were corrected for Faraday rotation using the derived RM map to obtain the intrinsic projected electric field vectors. These were rotated by 90\D\ to show the orientation of vectors of the magnetic field~(Figure~\ref{fig:fig7-rm}c). The vector lengths are scaled in proportion to the fractional polarisation~(Figure~\ref{fig:fig6-polarisation}c). The resulting projected magnetic field is highly ordered and aligned with the cometary morphology and tail direction of the nebula.

The polarisation analysis was repeated for \ah, yielding the same overall negative RM gradient observed at \al, although with significantly higher noise due to the reduced $\lambda^2$ leverage at high frequency. After derotation, the polarisation angle maps at both frequencies show consistent orientations, confirming a stable intrinsic magnetic field geometry across the nebula. Given that the \al\ band provides a more tightly constrained RM measurement, it is adopted for quantitative analysis, while the \ah\ results are used as a qualitative consistency check.

\subsection{X-ray properties of \pwnname}
\label{sec:res-x-ray}

In Figure~\ref{fig:fig8-x-ray}, we show the combined \xmm\ EPIC RGB image of \pwnname, revealing a hard X-ray counterpart coincident with the radio nebula. The X-ray source corresponds to 4XMM\,J163159.8$-$472156~\citep{2020A&A...641A.136W}, located at (R.A.,\,Dec)$_\text{J2000}$ = 16:31:59.8, $-$47:21:56. The source was detected with a net exposure of $\sim$23\,ks~(EPIC-pn, after vignetting correction) at relatively large off-axis angles of about 7\arcmin~(see~Table~\ref{tab:tab2-x-ray}). 

To investigate the presence of extended X-ray emission, we extracted a radial profile centred on the source in the 1$-$2\,keV energy range, where the signal-to-noise ratio is highest. To improve the statistics, we computed a weighted mean radial profile combining the individual EPIC-pn observations listed in Table~\ref{tab:tab2-x-ray}. The resulting profile was fitted with a Gaussian model, and the derived width was compared to that of the \xmm\ point spread function~(PSF) at the source position. 

\begin{figure}[!ht]
\includegraphics[width=\columnwidth]{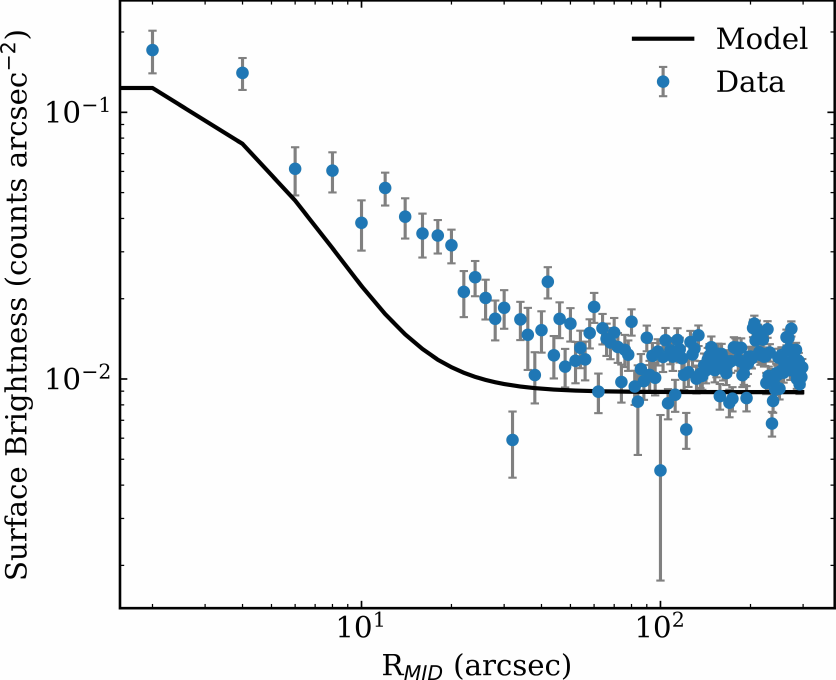}
\caption{Radial surface brightness profile of \pwnname\ PWN in the 1$-$2\,keV energy band.}
\label{fig:fig9-x-ray_spectar}
\end{figure}

\begin{figure}
\centering
\includegraphics[angle=-90, width=\columnwidth]{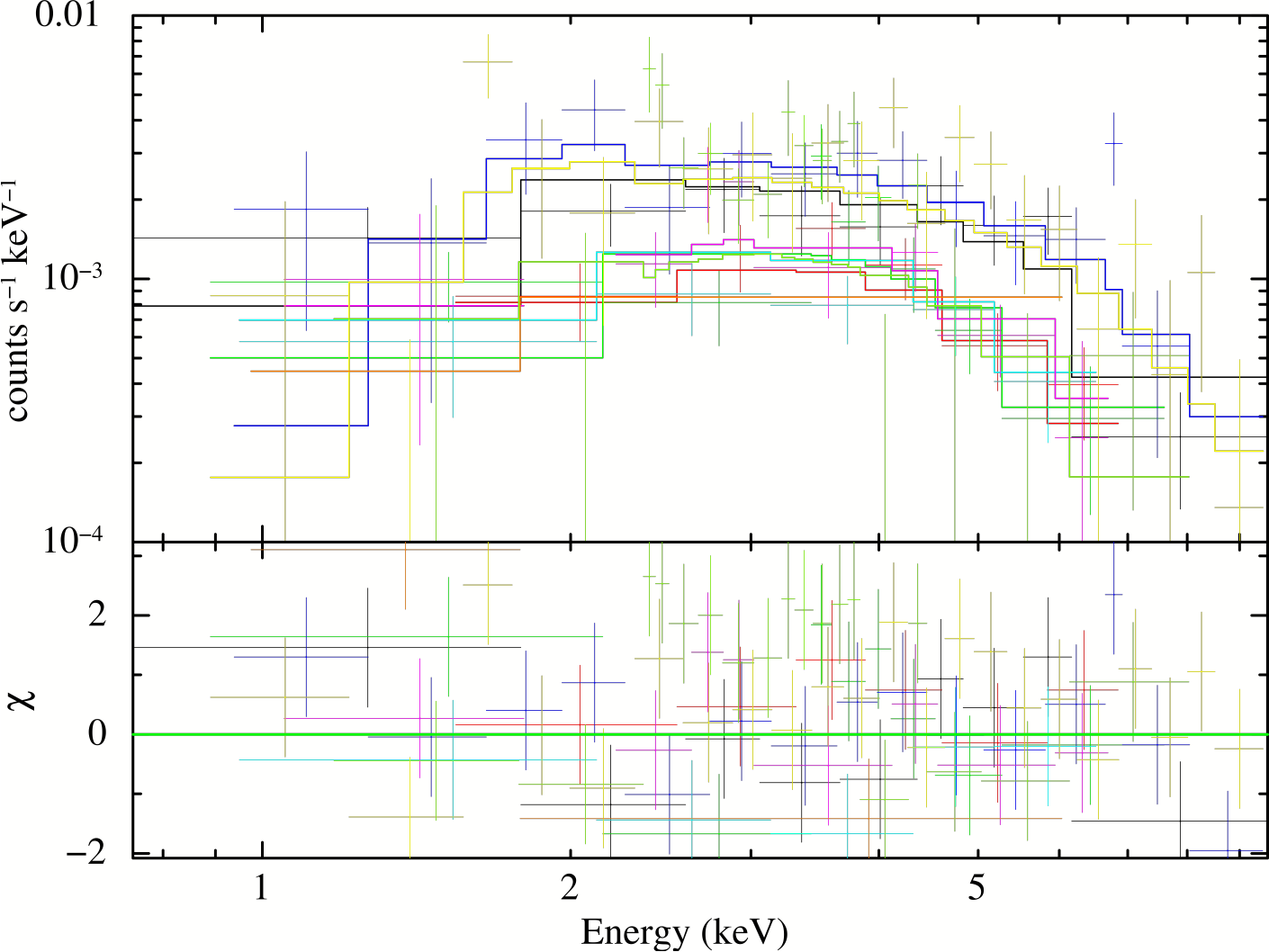}
\caption{The upper panel shows the simultaneous spectral fit using spectra from all available \xmm-EPIC cameras (Table~\ref{tab:tab2-x-ray}), together with the best-fit model described in Section~\ref{sec:res-x-ray}. The lower panel displays the residuals after the fit. The spectra have been rebinned for visual clarity.}
\label{figspec}
\end{figure}

In Figure~\ref{fig:fig9-x-ray_spectar}, we present the measured radial profile together with the PSF model. The detected X-ray emission extends to $\approx50$\arcsec, corresponding to a physical size of 1.7\,pc at a distance of 7\,kpc~(Section~\ref{sec:dis-distance}). This supports the interpretation of the emission as the X-ray counterpart of the radio PWN. The larger radio extent of $\simeq80$\arcsec\ is consistent with the longer cooling time of radio-emitting particles, which allows them to propagate further from the pulsar and produce a more extended nebular structure beyond the region of X-ray emission~\citep{2008AIPC..983..171K}. At the angular resolution of \xmm, the pulsar cannot be separated from the surrounding PWN emission. Future dedicated high angular resolution observations of \pwnname\ with \cxo\ will be required to resolve the pulsar from the nebular component. 

The X-ray spectrum of \pwnname\ was extracted and fitted simultaneously using all available observations. The spectrum is well described by an absorbed power-law model with a hydrogen column density of $N_{\rm H}$\,=\,3.08$^{+1.2}_{-0.91}\times10^{22}$\,cm$^{-2}$ and a photon index $\Gamma=1.6\pm0.4$. The estimated hydrogen column density $N_{\rm H}$ is significantly higher than the foreground Galactic value derived from the HI4PI survey\footnote{\href{https://heasarc.gsfc.nasa.gov/cgi-bin/Tools/w3nh/w3nh.pl}{HEASARC: Calculate the \hi\ Column Density}}~\citep{2016A&A...594A.116H}, indicating the presence of additional local absorption along the line of sight. 
The observed flux in the 0.2$-$12\,keV band is $2.3^{+0.2}_{-0.7} \times10^{-13}$\,erg\,cm$^{-2}$\,s$^{-1}$. 
Assuming a distance of 7\,kpc, this corresponds to an X-ray luminosity of 1.3\ergs{33}~(Table~\ref{tab:tab5-luminosity}). 
The best-fitting spectral model is shown in Figure~\ref{figspec}.

\section{Discussion}
\label{sec:discussion} 

Deep SKA pathfinder surveys resolve \pwnname\ as a cometary PWN emerging from the radio emission of its parent remnant, SNR\,\snr~(see Figure~\ref{fig:fig1-rgb}). Polarisation analysis reveals a highly ordered magnetic field aligned with the nebular flow, indicating efficient transport of relativistic particles along the tail. The detection of a hard X-ray counterpart further supports a synchrotron origin of the emission powered by the pulsar wind. Together with the radio properties of the SNR, these results suggest that the system represents an evolved stage, with \pwnname\ likely having escaped the remnant interior and now interacting with the surrounding ISM. 

Although the peaks of the radio and X-ray emission are spatially coincident~(see Figure~\ref{fig:fig8-x-ray}), the pulsar timing solution places \psr\ offset from this position. Given the relatively short timing baseline, as well as the presence of strong timing noise and a detected glitch, the derived timing position may still be affected by unmodelled rotational irregularities. Thus, the coincident radio and X-ray peaks likely provide a closer approximation to the true pulsar location.

\subsection{Location of \snrname--\pwnname\ in the Milky Way}
\label{sec:dis-distance}

Distance estimates to \snrname\ SNR have primarily relied on the empirical radio surface brightness–diameter~($\Sigma$\,$-$\,$D$) relation, yielding values of 6.3\,kpc~\citep{1970AuJPA..14..133S}, 8.7\,kpc~\citep{1998ApJ...504..761C}, 9.5\,kpc~\citep{2007Ap&SS.307..423S}, and 7.2\,kpc~\citep{2013ApJS..204....4P}. The most recent revision of the $\Sigma$\,$-$\,$D$ relation for Galactic shell-type SNRs gives distances of 6.1\,kpc and 6.9\,kpc using orthogonal-fit and probability-density approaches, respectively~\citep{2025SerAJ.211...45U}. 

An independent distance constraint is provided by the dispersion measure~(DM) of the powering pulsar \psr~\citep{2025MNRAS.537.2868A}. For DM\,$=$\,873\,pc\,cm$^{-3}$, the YMW16\footnote{\href{https://www.atnf.csiro.au/research/pulsar/ymw16/}{YMW16 electron density model}} Galactic electron density model~\citep{2017ApJ...835...29Y} yields a distance of 6.8\,kpc, while the NE2001 model~\citep{2002astro.ph..7156C} predicts a larger distance of 9.1\,kpc. 

Taken together, these estimates span approximately 6$-$9\,kpc. Although the results diverge along inner Galaxy sightlines, all methods constrain \snrname$-$ \pwnname\ system to a relatively narrow distance range, placing it within the Norma spiral arm~\citep{2007A&A...468..993K}. In addition, an examination of nearby pulsars~(see Figure~\ref{fig:fig11-psrs_rm_graph}) reveals a relative paucity of pulsars in the 8$-$9.5\,kpc distance interval, favouring distance estimates below 8\,kpc. We therefore adopt a fiducial distance of 7\,kpc throughout this work.

\subsection{Luminosity and surface brightness}
\label{sec:dis-luminosity}

Using the integrated radio spectral indices~(Table~\ref{tab:tab3-flux}), we extrapolate the flux densities to 1\,GHz, obtaining $S_{\rm 1GHz}$\,=\,5.6\,Jy for \snrname\ and 0.2\,Jy for \pwnname. From these values, we derive the corresponding monochromatic radio luminosities, radio powers and integrated radio luminosities over the frequency range 10\,MHz$-$10\,GHz~\citep{2006ARA&A..44...17G}. For \pwnname, we additionally report the X-ray luminosity of $L_{\rm x}$\,=\,1.34\,$\times10^{33}$\,erg\,s$^{-1}$, derived in the 0.2$-$12\,keV energy band~(Section~\ref{sec:res-x-ray}). For \snrname, we estimate the radio surface brightness at 1\,GHz, yielding $\Sigma_{\rm 1GHz}$\,=\,4.9\,$\times10^{-21}$\,W\,m$^{-2}$Hz$^{-1}$sr$^{-1}$ by adopting the elliptical region defined in Section~\ref{sec:res-morphology}. All derived quantities are summarised in Table~\ref{tab:tab5-luminosity}. 

The radio and X-ray luminosities of \pwnname\ place it well below those of young, Crab-like PWNe and firmly among more evolved systems~\citep{2018A&A...609A.110Z}. Since X-ray emitting electrons cool rapidly, whereas the radio emission trace the integrated history of particle injection, the ratio $L_{\rm X}$/$L_{\rm R}$ is expected to decrease with time and reach values of a few tens to a few hundreds. With a ratio of $\approx$130, \pwnname\ lies comfortably within the range expected for evolved PWNe. 

\begin{deluxetable}{l c c l}
\tablecaption{Radio flux densities, radio luminosities, X-ray flux and luminosity, and radio surface brightness of \snrname\ and \pwnname. A distance of 7\,kpc is assumed. All radio quantities are evaluated at 1\,GHz, except for the integrated radio luminosity, which is derived over the frequency range 10\,MHz$-$10\,GHz. The X-ray flux and luminosity are calculated in the 0.2$-$12\,keV energy band. Dashes indicate quantities that are not available or not applicable.}
\label{tab:tab5-luminosity}
\tablehead{
\colhead{} & \colhead{\snrname} & \colhead{\pwnname} & \colhead{Units}
}
\startdata
$S_{\rm 1GHz}$         & 5.6  & 0.2 & \small{[Jy]}\\
$L_{\rm \nu,1GHz}$     & 33.0 & 1.0 & \small{[10$^{22}$\,erg\,s$^{-1}$Hz$^{-1}$]}\\
$\nu L_{\rm \nu,1GHz}$ & 33.0 & 1.0 & \small{[10$^{31}$\,erg\,s$^{-1}$]}\\
$L_{\rm 10MHz-10GHz}$  & $-$  & 8.4 & \small{[10$^{31}$\,erg\,s$^{-1}$]}\\
$F_{\rm X}$            & $-$  & 2.3 & \small{[10$^{-13}$\,erg\,cm$^{-2}$ s$^{-1}$]}\\
$L_{\rm X}$            & $-$  & 1.3 & \small{[10$^{33}$\,erg\,s$^{-1}$]}\\
$\Sigma_{\rm 1GHz}$    & 4.9  & $-$ &\tiny{[10$^{-21}$\,W\,m$^{-2}$Hz$^{-1}$sr$^{-1}$]}\\
\enddata
\end{deluxetable}

We note that \citet{2025SerAJ.211...45U} report a surface brightness for \snrname\ of $\Sigma_{\rm 1GHz}$ = 6.4\,$\times10^{-21}$\,W\,m$^{-2}$Hz$^{-1}$sr$^{-1}$, which is consistent with our estimate. The small difference lies well within the uncertainties associated with defining the source boundary, background subtraction, and the scaling of the flux density to 1\,GHz. Modest variations in any of these factors can easily introduce differences at the 20$-$40\% level. In the context of the Galactic SNR population, \snrname\ falls within the main distribution of known remnants, following the surface brightness trends as reported by~\citet{2025ApJ...988...75B}, see their Figure~19. 
This contrasts with many newly discovered SNRs and SNR candidates, which typically exhibit significantly lower values, e.g., 1.4$\times10^{-22}$~W~m$^{-2}$~Hz$^{-1}$~sr$^{-1}$ for Ancora~\citep{2023AJ....166..149F}, 5.1$\times10^{-23}$~W~m$^{-2}$~Hz$^{-1}$~sr$^{-1}$ for Teleios~\citep{2025PASA...42..104F} and 2.4$\times10^{-22}$~W~m$^{-2}$~Hz$^{-1}$~sr$^{-1}$ for Abeona~\citep{2026arXiv260419897B}.

\subsection{Evolutionary status}
\label{sec:dis-evolutionaryStatus}


The radio morphology of \snrname\ and \pwnname\ indicates that the system is dynamically evolved. \snrname\ remnant displays a shell structure which the western side appearing significantly fainter and partially disrupted. Broken shell morphologies are commonly observed in evolved SNRs and are generally attributed to expansion into an inhomogeneous ISM~\citep{2017hsn..book.2211M,2019ApJ...884..113S}. In regions of lower ambient density, the shock propagates more rapidly, leading to shell thinning, fragmentation, and possible breakout, whereas denser regions confine the shock and enhance synchrotron emission.

The presence of the bow-shock PWN provides an independent evolutionary indicator. \pwnname\ pulsar was born in the same core-collapse event that produced \snrname\ SNR and likely received a substantial natal kick~\citep[][for review]{2004cetd.conf..276L}. 
Its supersonic motion through the ambient medium confines the pulsar wind via ram pressure, producing a bow-shock structure and a cometary nebula with a tail oriented opposite to the direction of motion. In moderately dense environments, the SNR reverse shock typically reaches the PWN on timescales of order 10$^4$\,yr~\citep{2001ApJ...563..806B,2001A&A...380..309V,2004AdSpR..33..475V}, further supporting an evolved stage of the system. 

No extended X-ray emission is detected from the SNR in the available archival data~(Section~\ref{sec:xmm}). This is consistent with an evolved remnant whose shock velocity has declined below the threshold required for efficient acceleration of electrons to X-ray emitting energies. While young SNRs with fast shocks~($\geq$1000\,km\,s$^{-1}$) commonly exhibit non-thermal X-ray shells, older remnants typically have shock speeds of only a few hundred km\,s$^{-1}$, shifting the synchrotron cutoff to lower energies and suppressing X-ray emission \citep{2006ApJ...648L..33V}. 
This effect is further compounded by the location of \snrname\ in the Norma region, where strong interstellar absorption~\citep{2001ApJ...547..792D} can efficiently obscure soft thermal X-ray emission. Therefore, the lack of detected X-rays does not exclude faint thermal plasma or weak non-thermal emission, but is consistent with an evolved SNR embedded in a highly obscured environment.

\subsubsection{Evolutionary models}
\label{sec:dis-evolutionarymodels}

To further constrain the evolutionary state of \snrname\ we derive upper limits on the thermal X-ray emission using the X-ray column density measured toward \pwnname~($N_{\rm H}$\,$\sim$\,1$\times10^{22}$\,cm$^{-2}$) and the upper limits on the SNR count rate from \xmm~(0.024\,counts\,s$^{-1}$, 0.2--12\,keV) and from SRG/eROSITA first-all-sky survey~(0.03\,counts\,s$^{-1}$, 0.2--1\,keV); for details see Section~\ref{sec:res-x-ray}. 
%
The \xmm\ limits were slightly more restrictive~(factor of $\sim$1.2), so we use those. Assuming an optically thin thermal plasma described by an APEC model with solar abundances, we obtained emission-measure~(EM) upper limits as a function of temperature~(Table~\ref{tab:emissionMeasure}). 

\begin{deluxetable}{c c c c c}[htp!]
\tablecaption{Upper limits on the EM for \snrname\ derived from \xmm-pn.
}
\label{tab:emissionMeasure}
\tablehead{
\colhead{T \lbrack K\rbrack} & \colhead{kT \lbrack keV\rbrack} & \colhead{EM$_{\mathrm{ul}}^{\textcolor{blue}{a}}$ \lbrack cm$^{-3}$\rbrack} & \colhead{Effect of absorption{\color{blue} $^{b}$}} 
}
\startdata
$4.0\times10^5$ & 0.034 & $1.65\times10^{61}$ & $2.38\times10^{-6}$ \\
$7.9\times10^5$ & 0.068 & $9.86\times10^{57}$ & $1.15\times10^{-4}$ \\
$2.0\times10^6$ & 0.172 & $2.87\times10^{55}$ & $8.54\times10^{-3}$ \\
$4.0\times10^6$ & 0.342 & $3.29\times10^{54}$ & $5.02\times10^{-2}$ \\
$1.0\times10^7$ & 0.861 & $1.01\times10^{54}$ & $1.41\times10^{-1}$ \\
\enddata
\tablecomments{{\color{blue}$^{a}$}Values are computed for the 0.2$-$12\,keV band. 
{\color{blue} $^{b}$}Ratio of the EM upper limit for $N_{\rm H}$\,=\,0 to that for $N_{\rm H}$\,=\,$1\times10^{22}$\,cm$^{-2}$.}
\end{deluxetable}

\begin{deluxetable}{c c c c c c c c c c c}
\tablecaption{Evolutionary models for \snrname. Models reproduce the observed mean radius at 7\,kpc. Columns list the ambient density, explosion energy, ejecta mass, SNR age, the transition time to the pressure-driven stage~(PDS), the reverse shock return time to the centre, forward- and reverse-shock temperatures, and EM.}
\label{tab:models}
\tablehead{\colhead{No.} & \colhead{ISM density{\color{blue} $^{a}$}} & \colhead{Energy{\color{blue} $^{a}$}} & \colhead{Ejected Mass{\color{blue} $^{a}$}} & \colhead{Age$_{\mathrm{SNR}}^{\textcolor{blue}{b}}$} & \colhead{Age$_{\mathrm{PDS}}^{\textcolor{blue}{b}}$} & \colhead{$t_{\mathrm{RS,centre}}^{\textcolor{blue}{b}}$} & \colhead{T$_{\mathrm{FS}}^{\textcolor{blue}{b}}$} & \colhead{T$_{\mathrm{RS}}^{\textcolor{blue}{b}}$} & \colhead{EM$^{\textcolor{blue}{b}}$} & \colhead{Consistent}\\ 
\colhead{ } & \colhead{\lbrack cm$^{-3}$\rbrack} & \colhead{\lbrack $10^{50}$\,erg\rbrack} &\colhead{\lbrack $M_{\odot}$\rbrack} & \colhead{\lbrack yr\rbrack} & \colhead{\lbrack yr\rbrack} & \colhead{\lbrack yr\rbrack} & \colhead{\lbrack $10^6$\,K\rbrack} & \colhead{\lbrack $10^6$\,K\rbrack} & \colhead{\lbrack $10^{60}$\,cm$^{-3}$\rbrack} 
}
\startdata
1  & 0.23 & 10 & 5   & 6200   & 31000 & 7300  & 18    & 15   & 2.4                          & N \\
2  & 0.23 & 10 & 20  & 7400   & 31000 & 23200 & 20    & 20   & 2.8                          & N \\
3  & 0.23 & 3  & 5   & 11300  & 24000 & 13300 & 7.4   & 6.4  & 2.4                          & N \\
4  & 0.23 & 3  & 20  & 13500  & 24000 & 42300 & 7.4   & 6.4  & 6.1                          & N \\
5  & 0.23 & 1  & 5   & 19600  & 19000 & 23100 & 2.3   & 21   & 1.2{\color{blue} $^{c}$}     & N \\
6  & 0.23 & 1  & 20  & 24300  & 19000 & 73300 & 1.4   & 2.5  & 1.2{\color{blue} $^{c}$}     & N \\
7  & 1.0  & 10 & 5   & 12100  & 13300 & 4500  & 5.7   & NA   & 1.2{\color{blue} $^{c}$}     & N \\
8  & 1.0  & 10 & 20  & 15200  & 13300 & 14200 & 3.8   & NA   & 1.4{\color{blue} $^{c}$}     & N \\
9  & 1.0  & 3  & 5   & 24500  & 10300 & 8200  & 0.94  & NA   & 0.57{\color{blue} $^{c}$}    & N \\
10 & 1.0  & 3  & 20  & 27500  & 10300 & 25900 & 0.79  & NA   & 0.46{\color{blue} $^{c}$}    & N \\
11 & 1.0  & 1  & 5   & 53300  & 8100  & 14100 & 0.18  & NA   & 0.036{\color{blue} $^{c}$}   & Y \\
12 & 1.0  & 1  & 20  & 66500  & 8100  & 44900 & 0.13  & NA   & 0.015{\color{blue} $^{c}$}   & Y \\
13 & 1.4  & 10 & 5   & 14400  & 11000 & 4000  & 3.3   & NA   & 1.0{\color{blue} $^{c}$}     & N \\
14 & 1.4  & 10 & 20  & 15300  & 11000 & 12700 & 3.0   & NA   & 0.95{\color{blue} $^{c}$}    & N \\
15 & 1.4  & 3  & 5   & 31200  & 8500  & 7300  & 0.53  & NA   & 0.089{\color{blue} $^{c}$}   & Y \\
16 & 1.4  & 3  & 20  & 43500  & 8500  & 23100 & 0.33  & NA   & 0.12{\color{blue} $^{c}$}    & Y \\
17 & 1.4  & 1  & 5   & 69800  & 6700  & 12600 & 0.10  & NA   & 0.003{\color{blue} $^{c}$}   & Y \\
18 & 1.4  & 1  & 20  & 87000  & 6700  & 40100 & 0.070 & NA   & 0.0006{\color{blue} $^{c}$}  & Y \\
\enddata
\tablecomments{{\color{blue}$^{a}$}Model inputs. {\color{blue} $^{b}$}Model outputs. {\color{blue} $^{c}$}Adjusted to account for cooling in the PDS stage.}
\end{deluxetable}

We compare the observational limits with SNR evolution models 
developed by \citet{2019AJ....158..149L}, including the EM calculations of \citet{2017AJ....153..239L}. The models are constrained to reproduce the observed mean radius of 13.4\,pc at a distance of 7\,kpc. The ambient density is estimated from the X-ray column density and distance, yielding $n_{\rm H}$\,$\sim$1.4\,cm$^{-3}$, and compared with the value inferred from Galactic density model~\citep[$\sim$0.23\,cm$^{-3}$;][]{2006A&A...459..113M}. In Table~\ref{tab:models}, we list the range of models and their properties, spanning the SNR ages from $\sim$\,6$-$90\,kyr depending on the assumed explosion energy and ambient density.

We computed models for three explosion energies: a typical explosion energy~($3\times10^{50}$\,erg), a low energy~($1\times10^{50}$\,erg) and a high energy~($1\times10^{51}$\,erg), consistent with values inferred for Galactic and Large Magellanic Cloud SNRs~\citep{2020ApJS..248...16L,2017ApJ...837...36L}. Given that the explosion produced the neutron star, we adopt ejecta masses of 5 and 20\,$M_{\odot}$, consistent with expectations for a core-collapse SNR. 
Models assuming low ambient density ($n_{\rm H}$\,=\,0.23\,cm$^{-3}$) predict plasma temperatures and EM values that exceed the observational X-ray limits~(Table~\ref{tab:models}). %
Acceptable solutions are obtained only for denser environments~($n_{\rm H}$\,$\gtrsim$\,1\,cm$^{-3}$) with shock temperatures below 6$\times$10$^5$\,K~(six of the last eight rows), in which the reverse shock has already reached the centre of the remnant. 

The viable models correspond to SNR ages of $>$\,30\,kyr, placing \snrname\ in the late Sedov phase approaching the transition to the pressure-driven snowplow stage~(PDS). This estimate can be independently compared with the characteristic age of the powering pulsar. 
According to canonical pulsar spin-down models, \citet{2025MNRAS.537.2868A} derived a characteristic age of $\tau_{\rm c}$\,=\,34\,kyr for \psr. The characteristic age can differ from the true pulsar age by a factor of a few~\citep[e.g.,][]{2002ApJ...567L.141M}, depending on the pulsar’s initial spin period and braking index~
\citep{2005ApJ...633.1095L}. Taken together with the radio morphology, these results suggest that \snrname--\pwnname\ system is best described by models 15 and 16 from Table~\ref{tab:models}, implying an age range of 30--45\,kyr.

\subsection{Radio spectro-polarimetry}
\label{sec:des-spectroPolimertry}

The radio spectral indices derived for \snrname\ and \pwnname~(Table~\ref{tab:tab4-spectralIndex}) provide clear evidence that the two sources arise from different particle acceleration environments. The spectral index measured for \snrname\ is consistent with expectations for shell-type SNRs, in which relativistic particles are accelerated at expanding shock fronts through diffusive shock acceleration~\citep[DSA;][]{Bell1978}. DSA theory predicts a characteristic radio spectral index of $\alpha$\,$\simeq$\,$-$0.5, which is in good agreement with observational studies, e.g.,~\citet{2023ApJS..265...53R,2024PASA...41..112F,2025A&A...693L..15S,2025PASA...42...69A}, and with typical variations of $\alpha$\,=\,$-$0.5\,$\pm$\,0.2~\citep{2017hsn..book.2041D}. 
In contrast, \pwnname\ exhibits a flatter spectrum characteristic of pulsar-powered synchrotron emission, produced by ultra-relativistic particles continuously injected by a rapidly rotating neutron star. Observed radio spectral indices for PWNe typically fall in the range $-$0.3\,$<$\,$\alpha$\,$<$\,0~\citep{2012SSRv..166..231R,2017ASSL..446....1K}, fully consistent with the observed spectral behaviour of \pwnname. 
The clear spectral distinction between \snrname\ and \pwnname, combined with their morphologies, strongly supports a scenario in which \pwnname\ represents a pulsar-powered nebula projected within the SNR shell of \snrname. 

A flat radio spectrum alone is not sufficient to uniquely identify synchrotron emission, as optically thin thermal emission from \hii\ regions can produce similar spectral slopes. However, the emission mechanisms of these systems differ fundamentally, where synchrotron emission is intrinsically linearly polarised, whereas thermal free–free emission is effectively unpolarised~\citep{2005ism..book.....L}, making polarimetry a powerful diagnostic for distinguishing between thermal and non-thermal radio nebulae.  

PWNe typically show fractional linear polarisation ranging from a few per cent up to $\gtrsim$\,30\%, depending on magnetic-field coherence and the degree of Faraday depolarisation~\citep{2022hxga.book...61M}. \pwnname\ shows significant linear polarisation across the nebula, with average fractional polarisation of $\sim$10\% at \al\ and $\sim$18\% at \ah. These values indicate that the magnetic field is 
ordered on spatial scales larger than the beam.

The polarised intensity distribution of \pwnname\ does not closely follow the total intensity morphology. The peak of Stokes\,$I$ emission coincides with a local minimum in fractional polarisation~(Figure~\ref{fig:fig6-polarisation}), suggesting enhanced depolarisation and/or increased magnetic field disorder in this region. This behaviour is consistent with enhanced turbulence, higher local electron densities, or stronger line-of-sight Faraday rotation where the pulsar wind interacts directly with the surrounding medium~\citep{2017ASSL..446....1K,2012SSRv..166..231R}. 
Higher fractional polarisation is observed in the fainter regions of the tail, reaching values of up to $\sim$30\%. While partially influenced by reduced total intensity, the simultaneous increase in polarised intensity suggests intrinsically higher magnetic field ordering and a more laminar outflow, similar to that observed in other PWNe~\citep[e.g.,][]{2010ApJ...712..596N,2006ApJ...638..225K}.

The systematically higher fractional polarisation observed at \ah\ relative to \al\ is consistent with synchrotron emission propagating through a magneto-ionic medium, where Faraday rotation produces stronger depolarisation at longer wavelengths~\citep{2021MNRAS.507.4968F}. Such frequency-dependent depolarisation is commonly observed in PWNe and other synchrotron-dominated systems~\citep[e.g.,][]{2012ApJ...746..105N,2020MNRAS.496..723K,2023MNRAS.523.1933V,2024MNRAS.534.2918S}.

\subsection{Faraday environments}
\label{sec:dis-rm}

The RM map derived from the ATCA \al\ data~(Figure~\ref{fig:fig7-rm}a) reveals uniformly negative Faraday depth across \pwnname, with an average value of $-752$\,rad\,m$^{-2}$ and a typical uncertainty of 31\,rad\,m$^{-2}$. The consistent sign of the RM across the nebula 
suggests an ordered Faraday-rotating medium along the line of sight. 

\begin{figure}
\centering
\includegraphics[width=1\columnwidth]{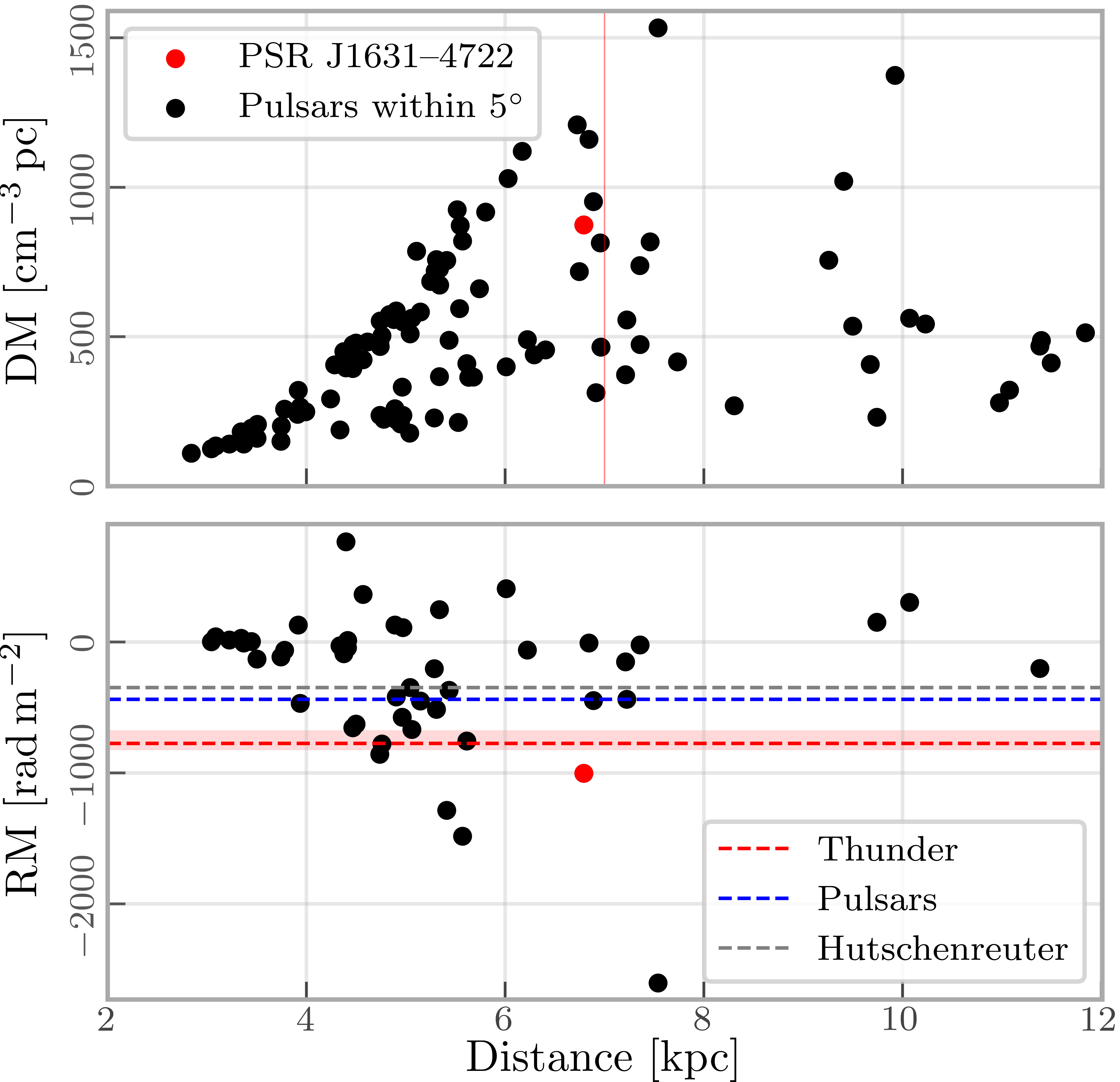}
\caption{DM~(top) and RM~(bottom) as a function of distance for the pulsar population located within 5\D\ of \psr, which is highlighted in red. The vertical red line marks the adopted distance of 7\,kpc. In the lower panel, the blue dashed line indicates the median RM of the surrounding pulsars, while the grey dashed line represents the median Galactic foreground RM toward \snrname\, taken from the model of \citet{2022A&A...657A..43H}. The red dashed line and the shaded region show the median and RM range measured for \pwnname~(Section~\ref{sec:res-rm}).}
\label{fig:fig11-psrs_rm_graph}
\end{figure}

To estimate the Galactic foreground contribution, we consider two approaches. We use the all-sky Galactic RM model\footnote{\href{https://wwwmpa.mpa-garching.mpg.de/~ensslin/research/data/faraday2020.html}{The Galactic Faraday rotation sky 2020}} developed by \citet{2022A&A...657A..43H}. For an aperture of the SNR angular size, the model predicts a median foreground RM of $-$348\,$\pm$\,171\,rad\,m$^{-2}$, with only weak spatial variation across the region~($\simeq$\,27\,rad\,m$^{-2}$). We also examine the RMs of pulsars located within 5\D\ of \psr~(Figure~\ref{fig:fig11-psrs_rm_graph}), using the ATNF Pulsar Catalogue\footnote{\href{https://www.atnf.csiro.au/research/pulsar/psrcat/}{ATNF Pulsar Catalogue}}~\citep{2005AJ....129.1993M}. Restricting the sample to pulsars within $\pm$1.5\,kpc of \psr\ position, we find a median RM of $-$438\,rad\,m$^{-2}$. These two approaches probe complementary aspects of the magneto-ionic medium, where the model provides a smoothed large-scale estimate of the Galactic foreground, while pulsars sample the integrated Faraday rotation along discrete nearby lines of sight.

Comparing these foreground estimates with the observed RM of \pwnname\ suggests an excess rotation of roughly $-$300 to $-$400\,rad\,m$^{-2}$. Such an excess may arise from magneto-ionic material local to the nebula, although uncertainties in the foreground structure and the limited sampling of nearby pulsars could lead to an underestimate of the true Galactic contribution.

PWNe are dominated by relativistic plasma, contain relatively little thermal material, and are generally expected to be internally Faraday thin, contributing negligibly to Faraday rotation~\citep{2012SSRv..166..231R}. In this case, the Faraday-active medium is expected to lie external to the synchrotron-emitting plasma but local to the system, most plausibly originating from mixing between nebular material and supernova ejecta that has been heated and compressed by the reverse shock. 

Using the standard relation RM\,=\,0.81$\int n_e\,B_{\mid\mid}\,dl$, and adopting post-shock electron densities of $n_e$\,$\approx$\,10$-$25\,cm$^{-3}$~\citep{2001ApJ...548..820B,2019MNRAS.484.4760B}, together with a characteristic Faraday path length of $l\,\simeq\,$0.5\,pc~\citep[corresponds to an order-unity fraction of the compact nebula width;][]{2019MNRAS.484.4760B,2019MNRAS.488.5690O}, the observed RM excess implies a line-of-sight magnetic field strength of $B_{\mid\mid}$\,$\approx$\,$-$30 to\,$-100\,\mu$G. The negative sign indicates that the magnetic field is directed away from the observer, consistent with the uniformly negative RM observed across the nebula. Field strengths of this magnitude are typical of compressed magnetic fields in shocked environments and compatible with an origin of the Faraday rotation within the bow-shock sheath~\citep{2012SSRv..166..231R}. 

The RM map also reveals a smooth gradient across \pwnname~(Figure~\ref{fig:fig7-rm}a). 
Such behaviour may indicate the presence of a Faraday screen associated with the nebula itself. One possible interpretation is a geometric effect. If the bow shock is tilted relative to the line of sight, the effective path length through the Faraday-active plasma would vary smoothly across the nebula, producing a monotonic RM gradient. Alternatively, variations in electron density or magnetic field strength in the surrounding medium could produce a similar effect. The present data do not allow these possibilities to be distinguished.

An additional constraint on the magneto-ionic environment comes from the pulsar itself. \psr\ probes 
the magneto-ionic medium integrated along the full line of sight. 
Using the measured values DM\,=\,874\,pc\,cm$^{-3}$ and RM\,=\,$-$1004\,rad\,m$^{-2}$~\citep{2025MNRAS.537.2868A}, the mean projected magnetic field is $\langle B_{\mid\mid}\rangle\,$=\,1.23\,RM/\,DM$\,\approx$\,$-$1.4\,$\mu$G. Adopting the distance of 7\,kpc gives a mean electron density of $\langle n_e\rangle\,$=\,DM/\,d$\,\simeq$\,0.13\,cm$^{-3}$, consistent with the Galactic electron-density model YMW16 
\citep{2017ApJ...835...29Y}. 
Because both RM and DM represent electron-weighted integrals along the entire propagation path, these quantities reflect the large-scale Galactic magneto-ionic medium. 
The pulsar RM is larger in magnitude than both the median RM measured for nearby pulsars and the foreground value predicted by the Galactic RM model. Although this difference may indicate additional Faraday rotation arising in the local environment of the system, uncertainties in the Galactic foreground structure and sparse sampling of pulsars in this region prevent a unique interpretation. In this context, the enhanced RM observed across the PWN may arise from locally compressed magneto-ionic plasma associated with the bow-shock region. 

\subsection{Magnetic field geometry and strength of \pwnname}
\label{sec:dis-mf}

The polarimetric properties of \pwnname\ reveal a highly ordered magnetic field 
shaped by the nebular flow~(see Figure~\ref{fig:fig7-rm}c). The projected magnetic field vectors are aligned with the symmetry axis of the radio emission, 
tracing the elongated morphology of the nebula. Similar magnetic field alignment has been observed in several other cometary PWNe, including the tail of the Mouse~\citep{2005AdSpR..35.1129Y}, the handle of the Frying pan~\citep{2012ApJ...746..105N}, the head of the Snail~\citep{2016ApJ...820..100M}, the Goose~\citep{2022ApJ...932...89K} and the Zig-Zag PWN~\citep{2025SerAJ.211...27G}. 

Magnetohydrodynamic simulations predict that, in such systems, the magnetic field is compressed and stretched into an elongated magnetotail, particularly for highly magnetised pulsar winds~\citep{2019MNRAS.484.5755O}. In this regime, particle transport is dominated by advection along ordered magnetic field lines, while cross-field diffusion is suppressed. This configuration forms a magnetic flux tube-like structure that efficiently channels relativistic particles downstream. The resulting flow aligned magnetic geometry shapes the PWN morphology and can produce anisotropic pitch-angle distributions, causing the synchrotron emission to become beamed and dependent on the relative orientation between the magnetic field, flow direction, and the observer’s line of sight~\citep{2024NatAs...8.1284K}. 

To estimate the magnetic-field strength of \pwnname, we assume energy equipartition between relativistic particles and the magnetic field. Following the formalism of
\citet{1970ranp.book.....P}, the equipartition magnetic field is B$_{\rm eq}$\,=\,$[6\,\pi\,c_{12}\,(1$\,+\,$k)\,L\,\Phi^{-1}\,V^{-1}]^{2/7}$, where $L$ is the integrated synchrotron luminosity, $V$ is the emitting volume, $k$ is the ion-to-electron energy ratio, $\Phi$ is the volume filling factor, and $c_{12}$ depends weakly on the spectral index and the adopted frequency range. 
This classical equipartition formulation provides an order-of-magnitude estimate of the nebular magnetic-field strength and is widely used for synchrotron sources such as PWNe. 

For a synchrotron spectrum integrated over 10$^7$$-$10$^{11}$\,Hz, assuming $k$\,=\,0 and $\Phi$\,=\,1, we derive an equipartition magnetic field strength of B$_{\rm eq}$\,$\approx$\,54\,$\mu$G. Extending the integration to 10$^{13}$\,Hz, assuming no spectral break, increases the estimate to B$_{\rm eq}$\,$\approx$\,140\,$\mu$G, which we treat as an upper limit. The inferred field strength is largely insensitive to the choice of the lower-frequency cutoff. 
The estimate is subject to systematic uncertainties related to the assumed emitting volume, filling factor, and particle composition, but these do not affect the order-of-magnitude field strength inferred for the nebula. 
These values are consistent with magnetic-field strengths commonly inferred in PWNe~($\sim$\,1$-$100\,$\mu$G), and with enhanced fields expected in compressed pulsar winds and bow-shock PWNe~\citep[e.g.,][]{2017SSRv..207..175R,2025OJAp....854247G}. The ordered field geometry and equipartition field strength support a scenario in which the PWN magnetic field is amplified and aligned by ram pressure confinement in the bow-shock flow downstream of the pulsar. 

\subsection{Comparison of radio and X-ray properties of \pwnname}
\label{sec:dis-comparison}

The radio and X-ray emission from \pwnname\ 
trace different populations of relativistic particles injected by \psr\ and probe distinct cooling regimes within the nebula. The X-ray nebula is more compact, extending to $\sim$50\arcsec, compared to the radio nebula, which extends to $\sim$80\arcsec. This difference is explained by synchrotron ageing of the highest-energy electrons. Particles emitting in X-rays cool rapidly and remain confined close to the pulsar, while lower-energy electrons responsible for the radio emission survive longer and populate a larger downstream volume~\citep[e.g.,][]{2014A&A...562A.122P,2018ApJ...861....5K}.


The radio and X-ray spectral properties further support a synchrotron origin 
for both components. The radio emission follows a power-law spectrum with spectral index $\alpha = -0.27 \pm 0.05$, consistent with the spectral distribution observed in known PWNe~\citep{2017ASSL..446....1K}. The X-ray emission is also well described by a power-law photon spectrum with photon index $\Gamma = 1.6 \pm 0.4$. 
%
For PWNe, photon indices are $\Gamma$\,$\sim$\,2~\citep{2006ARA&A..44...17G}, placing \pwnname\ on the harder side of the PWNe distribution. Within uncertainties, the radio and X-ray spectral values are broadly consistent with synchrotron radiation produced by a common population of relativistic electrons, with the radio emission tracing lower-energy particles and the X-rays probing the highest-energy electrons close to the pulsar. 

For the equipartition magnetic field inferred from the radio emission~($B_{\rm eq} \approx 54$–$140\,\mu$G; Section~\ref{sec:dis-mf}), high-energy electrons are expected to cool rapidly via synchrotron losses. The synchrotron cooling time of electrons emitting photons of energy $E_{keV}$ in a magnetic field $B$ is $t_{syn}$\,$\approx$\,1.1\,$\times$\,10$^3$ (B/10\,$\mu$G$)^{-3/2} (E_{\rm keV}/1)^{-1/2}$\,yr~\citep{2008ApJ...684..542K}. Using the inferred equipartition magnetic field, electrons producing 1$-$5\,keV X-ray photons cool on timescales of order $t_{syn}$\,$\sim$100\,yr. 
In contrast, electrons radiating at radio frequencies have cooling times of order $t_{syn}\approx 10^6$\,yr and therefore remain largely unaffected by radiative losses over the flow lifetime of the nebula. These cooling times naturally explain the smaller extent of the X-ray emission relative to the radio nebula and support a flow-dominated bow-shock PWN morphology. 
%

While comparisons of PWN properties have largely focused on radio and X-ray imaging and spectral analysis, polarimetric studies have traditionally relied on radio observations. More recently, observations with the Imaging X-ray Polarimetry Explorer \citep[IXPE;][]{2022JATIS...8b6002W} have extended such studies into the X-ray regime, providing complementary constraints on magnetic-field geometry, ordering, and particle transport~\citep[e.g., Lighthouse PWN;][]{2026arXiv260422914D}. In this context, the radio polarisation properties of \pwnname\ provide important information on the nebular magnetic-field structure and complement emerging high-energy polarimetric studies of PWNe.

\section{Conclusion}
\label{sec:conclusion}

We report the discovery of \pwnname, a bow-shock PWN powered by \psr\ and projected within SNR \snr~(\snrname). This study combines deep radio observations from ASKAP--EMU and MeerKAT with dedicated ATCA follow-up observations and archival \xmm\ data to investigate the structure and environment of the system.

The radio emission reveals an elongated cometary nebula with a highly ordered magnetic field aligned with the direction of the nebular flow and a coherent Faraday rotation gradient across the nebula. 
%
The radio spectrum, polarisation structure, and magnetic field geometry indicate efficient downstream transport of relativistic particles and the preservation of ordered magnetic fields along the full extent of the tail. An equipartition estimate yields a nebular magnetic-field strength of B$_{\rm eq}$\,$\approx$\,54$-$140\,$\mu$G, consistent with values typically inferred for bow-shock PWNe. 

We also identify X-ray 
emission coincident with \pwnname\ in archival \xmm\ observations. The X-ray spectrum is consistent with synchrotron radiation from relativistic particles accelerated in the pulsar wind. 
In contrast, \snrname\ shows no convincing diffuse 
X-ray emission in the available data, consistent with an evolved remnant expanding into a relatively tenuous environment and/or significant absorption along the line of sight. 

Taken together, the radio and X-ray results support an evolutionary scenario in which the pulsar has travelled from its birth site near the geometric centre of \snrname\ and is now interacting with the surrounding medium, producing the observed bow-shock magnetotail. The morphology and multiwavelength properties of the \snrname--\pwnname\ system indicate that \snrname\ is an evolved SNR with an estimated age of 30$-$45\,kyr, placing Nimbus in the late Sedov phase approaching the transition to the radiative stage. 

\pwnname\ and \snrname\ provide a valuable observational link between PWN evolution, pulsar escape, and the large-scale environment of the SNR. Systems in this transitional stage remain relatively rare, yet offer important insight into particle transport, magnetic-field evolution, and the late-time interaction between neutron stars and their birth remnants. Future deep X-ray observations, high-resolution broadband radio imaging, and continued pulsar timing will be crucial for constraining particle energetics, identifying spectral breaks, and refining the pulsar timing solution, thereby further probing the evolutionary history of this system. 
More broadly, the discovery of \pwnname\ highlights the power of SKA precursor surveys to reveal new physical insight within both newly discovered and previously known Galactic objects.

\begin{acknowledgments}
This scientific work uses data obtained from Inyarrimanha Ilgari Bundara, the CSIRO Murchison Radio-astronomy Observatory. We acknowledge the Wajarri Yamaji People as the Traditional Owners and native title holders of the Observatory site. CSIRO’s ASKAP radio telescope is part of the Australia Telescope National Facility~(ATNF; \href{https://ror.org/05qajvd42}{https://ror.org/05qajvd42}). Operation of ASKAP is funded by the Australian Government with support from the National Collaborative Research Infrastructure Strategy. ASKAP uses the resources of the Pawsey Supercomputing Research Centre. Establishment of ASKAP, Inyarrimanha Ilgari Bundara, the CSIRO Murchison Radio-astronomy Observatory and the Pawsey Supercomputing Research Centre are initiatives of the Australian Government, with support from the Government of Western Australia and the Science and Industry Endowment Fund. 
Murriyang, CSIRO’s Parkes radio telescope, is part of the ATNF. We acknowledge the Wiradjuri people as the Traditional Owners of the Observatory site. 

This research has made use of NASA’s Astrophysics Data System~\citep{2000A&AS..143...41K} and the SIMBAD database~\citep{2000A&AS..143....9W}. We have used software Astropy~\citep{2013A&A...558A..33A}, DS9~\citep{2003ASPC..295..489J}, CARTA~\citep{carta} and Aladin~\citep{2000A&AS..143...33B} at various stages of this research. 
This research has made use of data obtained from the 4XMM XMM-Newton Serendipitous Source Catalog compiled by the 10 institutes of the XMM-Newton Survey Science Centre selected by ESA.

S.L., M.D.F., and G.R. acknowledge the Australian Research Council funding through grant DP200100784. 
S.H. acknowledges funding by the European Union (ERC, ISM-FLOW, 101055318). Views and opinions expressed are, however, those of the author(s) only and do not necessarily reflect those of the European Union or the European Research Council. Neither the European Union nor the granting authority can be held responsible for them.

\end{acknowledgments}





\bibliography{main}{}
\bibliographystyle{aasjournalv7}

\end{document}